\documentclass[amsmath,amssymb,aps,pre,preprint,groupedaddress]{revtex4}
\usepackage{epsfig,setspace,graphics,amsmath}

\input xy
\xyoption{all}

\newcommand{\bra}[1]{\langle #1 |}
\newcommand{\ket}[1]{| #1 \rangle}

\begin{document}

\title{ Single molecule photon counting statistics for quantum mechanical
chromophore dynamics }
\author{Golan Bel, Yujun Zheng and Frank L. H. Brown}
\affiliation{Department of Chemistry and Biochemistry, 
University of California, Santa Barbara, CA 93106-9510}

\begin{abstract} 
We extend the generating function technique for calculation of single molecule
photon emission statistics [Y. Zheng and F. L. H. Brown, Phys. Rev. Lett.,
90,238305 (2003)] to systems governed by multi-level quantum dynamics.
This opens up the possibility to study phenomena that are outside the realm of
purely stochastic and mixed quantum-stochastic models.  In particular, the present
methodology allows for calculation of photon statistics that are spectrally
resolved and subject to quantum coherence.  Several model 
calculations illustrate the generality of the technique and highlight quantitative 
and qualitative differences between quantum mechanical models and related 
stochastic approximations.  Calculations suggest that studying photon statistics
as a function of photon frequency has the potential to reveal more about system
dynamics than the usual broadband detection schemes.
\end{abstract}

\maketitle

\section{introduction}
Single molecule spectroscopy (SMS) has become
a versatile and powerful tool for the study of condensed phase systems in 
chemistry, physics and biology \cite{moerner,orrit,moernerrev,
wildrev,xie,xie_rev,chu,sweiss,silbrev,speculate,cpc_rev,orrit_rev}.  Unfortunately, the
very qualities that make SMS such a powerful technique, have
also led to significant theoretical challenges in describing 
experimental data.  The ultra-microscopic nature of the 
physical systems under study leads to randomness in the behavior
of experimental signals due both to thermal agitation of 
the photoactive portion of the system and the inherent randomness
of the spontaneous emission process itself.  
While SMS 
has been hailed for its ability to probe these fluctuations directly, it remains difficult to extract
physical pictures for molecular dynamics based solely on SMS data streams.  Some of this
difficulty is likely fundamental (current SMS experiments may not collect sufficient data
to allow for direct inversion to molecular dynamics), but even if SMS data were sufficient to
differentiate between all viable physical hypotheses, it remains an open question as
to the best means to simulate such models to allow for comparison with experiment.
Indeed, much effort has been expended on the theory of interpreting/modeling SMS trajectories,
particularly in the context of stochastic models for chromophore dynamics
\cite{silbey_rev,orrit_rev,skinner,wang,wangII, schenter,portman,agmon,cao,caoII,mukamel,
muk3,barkai,younjoon,brown,wild,szabo1,szabo2,orphir}.  Stochastic models, though certainly
illustrative and powerful, ultimately face certain limitations in the modeling of phenomena
that are inherently quantum mechanical, such as spectroscopy.  Quantum coherence can 
not be captured, quantization of nuclear eigenstates is not naturally formulated within a stochastic
scheme, and the parameters of stochastic models are often difficult to equate with their microscopic 
origins.   As the following work will show, even stochastic models systematically derived from 
underlying quantum considerations can lead to quantitative and qualitative differences from
fully quantum calculations.

Until recently, Monte Carlo Wave Function simulations (MCWF) \cite{plenio}
and related techniques provided the only fully satisfactory route to theoretical 
calculation of single molecule photon counting observables \cite{metiu} including
quantum mechanical effects.  A few
other studies have touched on specific aspects of quantum dynamics applicable to
SMS \cite{wild,osadko,wild2,barkai,barkrev}, but without complete 
generality.  Recent work by us \cite{zheng1,zheng2,zheng3,zheng4} and others 
\cite{cook,muk1,bark1,bark2,muk2},has established generating function 
techniques as a general means for calculating statistical quantities of single molecule
photon counting experiments.  The only fundamental limitations to this approach are
that you must consider the spontaneous emission of photons to be governed by
rate processes  and the directly calculated quantities 
are statistical moments of the number of photons emitted \cite{zheng1,zheng2,muk1}.

The bulk of previous work with the generating function approach has focused on two
level chromophores with stochastic modulation by the environment, however the 
method is equally applicable to multi-state quantum systems.  
The extension to multi-state quantum systems was suggested by us
\cite{zheng2} and formally carried out by Mukamel \cite{muk1}.  Sanda and Mukamel 
\cite{muk2} have recently used the generating function approach to derive formal
perturbative expressions (in the applied field strength) for low order photon counting 
moments.  Though interesting from a theoretical standpoint, the derived expressions
are complex enough that implementation will be impossible for all but the simplest model
systems (second order moments require solution of a six point quantum
correlation function, higher moments need larger correlations).  As a numerical technique,
the generating function approach has promise to study varied systems without limitation
to low field strengths.

The present paper considers several model systems to demonstrate
the use of the generating function approach as a numerical tool for predicting
SMS photon counting observables.  In addition to 
calculation of photon counting moments for broadband detection schemes, 
as has been considered previously, we also calculate emission statistics for 
photons specific to particular molecular transitions and degenerate sets of transitions. For 
systems where vibrational structure is well resolved compared to natural line widths, this
is equivalent to the calculation of spectrally resolved emission statistics.  From a conceptual
and numerical standpoint these calculations are no more difficult than broadband detection
calculations.  The simulations we have carried out suggest that significantly more information
stands to be learned from photon counting experiments when photon statistics are broken
down by color.

This paper is organized as follows.  Section \ref{sec:theory} presents the underlying theory
and notation necessary to introduce our calculations.  Although there are many details to
be considered here, the conceptual framework for calculating photon statistics in the many-level
case is no more complex than for two level chromophores.  Given the reduced Liouvillian operator
for density matrix dynamics of the chromophore system, calculation of the generating function 
for photon number and/or low order statistical moments is immediate.  Most of sec. \ref{sec:theory}
is dedicated to describing the Liouvillian operator itself, not the extension of this matrix
to calculation of experimental observables.  Sections \ref{sec:tls} and \ref{sec:harmonic}
present numerical calculations for chromophores coupled to a two 
level system and an harmonic vibrational coordinate.  Many different regimes are considered, both
to display the flexibility of the present formulation in numerical calculations 
and to highlight differences between fully quantum calculations and commonly 
employed stochastic approximations.  In sec. \ref{sec:conc} we conclude.

\section{Theoretical background}
\label{sec:theory}
\subsection{General considerations for chromophore dynamics}
The picture we present is the natural extension of the optical 
Bloch equations to multi-level quantum systems in a condensed
phase.  Our methodology has been adopted both to make connection
with our previous work on two level chromophores \cite{zheng1,zheng2,zheng3}
and because the necessary theoretical/computational tools for 
chromophore dynamics are well established in the literature.

We imagine a single chromophore in a condensed phase
environment driven by an external 
laser field.  It is assumed that the field is strong enough
to warrant a classical treatment of this perturbation so that
dynamics, in the absence of any other system-field interactions,
would be dictated by
\begin{equation}
\label{eq:no_relax}
\dot{\rho}(t)= -\frac{i}{\hbar}[\hat{H}^{sys},\rho] + \frac{i}{\hbar}\boldsymbol{E}(t)\cdot[ \hat{\boldsymbol{\mu}},\rho].
\end{equation}
In the above $\hat{H^{sys}}$ is the Hamiltonian for the unperturbed
chromophore-environment system, $\hat{\boldsymbol{\mu}}$
is the electric dipole moment for this system and $\boldsymbol{E}(t)$ is the classical
applied laser field.  $\rho(t)$ specifies the density matrix for the molecule only.
This evolution assumes the radiation is of sufficiently long wavelength (and the
chromophore sufficiently localized) to allow the dipole approximation.  

What the above dynamics neglects is the relaxation 
of the driven molecular system.  The coupling between system and the quantum
radiation field provides a route for this relaxation to occur: the spontaneous
emission of photons.  It is these photons that are registered in SMS experiments and hence
inclusion of the spontaneous emission process is absolutely essential.  Within the 
standard approximations, the quantum radiation field is integrated over to provide
rate constants for emission of photons between various molecular transitions
\cite{loudon,ct}.  This leads to a master equation approach for incorporating
emission events as pure rate processes.
The rate for spontaneous emission of a photon, causing a
jump from system eigenstate $i$ to eigenstate $j$, is calculated by application of Fermi's
golden rule (using the coupling between system and quantum radiation field as 
the perturbation)
\begin{eqnarray}
\label{eq:emit_rate}
\Gamma_{ij} &=& \frac{ \omega_{ij}^3 D_{ij}^2}{3\pi \epsilon \hbar c^3} \\ \nonumber
\boldsymbol{D}_{ij} &=& \bra{i}\hat{\boldsymbol{\mu}}\ket{j}.
\end{eqnarray}
The collection of constants appearing in this expression have their usual meaning, but
we will not be concerned with them in this work.  What is important to us is 
the dependence on the transition dipole moment $\boldsymbol{D}_{ij}$, which serves 
to mediate relative rates of emission for different chromophore transitions.  
In principle, energy splittings ($\omega_{ij}$) impact the rates as well, 
but we shall be concerned with 
electronic transitions where differences in this quantity between various allowed transitions 
are much smaller than the splitting itself.  In this limit we expect inconsequential variations 
on the basis of energy differences.

Perturbation theory applied to the entire system density matrix evolution (as opposed to just a single
rate calculation) additionally tells us that the population lost from state $i$, via $\Gamma_{ij}$
decay, ends up in state $j$.  Also, it specifies that the $i  \rightarrow j$ transition causes all
associated coherences ($\rho_{ik}$, $\rho_{ki}$) to decay at the rate of $\Gamma_{ij}/2$.  The net effect of all spontaneous emission
processes in the system is the additive contribution of these three effects (loss of population from
state $i$, gain in population of state $j$ and loss of coherence for all allowed $i\rightarrow j$
transitions.)
We neglect radiative level
shifts in the system states and ignore all other couplings
(virtual photon transitions) caused by the presence of the quantum radiation field.  These
other couplings are unimportant when system energy levels are non-degenerate as the implied
perturbations are non-secular \cite{loudon,ct}.  The non-degeneracy condition is met by the
systems studied in this work.

Keeping those contributions specified in the last two paragraphs, implies that we supplement
our chromophore equations of motion with non-Hamiltonian evolution terms corresponding to
spontaneous emission.  The form of this augmentation is most transparent in the basis of $\hat{H}^{sys}$ 
eigenstates.  Rewriting eq.
\ref{eq:no_relax} in this form yields (The summation convention over repeated density matrix
labels is assumed throughout this work.)
\begin{equation}
\label{eq:with_q_field_rem}
\dot{\rho}_{ij}(t) = \mathbb{L}^{sys}_{ij;kl}\rho_{kl} +\mathbb{L}^{E}_{ij;kl}(t)\rho_{kl} +\mathbb{L}^{\Gamma}_{ij;kl}\rho_{kl}.
\end{equation}
Here, $\mathbb{L}^{sys}$ and $\mathbb{L}^E (t)$ are Liouville super-operators (matrices) corresponding to
the commutator expressions in eq. \ref{eq:no_relax} \cite{mukbook}.  Note that our definition incorporates
the factor $-i/\hbar$ within $\mathbb{L}^{sys}$ and $\mathbb{L}^E (t)$.   $\mathbb{L}^{\Gamma}$ is the matrix
effecting spontaneous emission processes.    The elements of this matrix are provided by the
arguments of the preceding paragraphs ($i \neq j$ assumed in the following)
\begin{eqnarray}
\label{eq:def_L}
\mathbb{L}^{\Gamma}_{ii;ii} &=& -\sum_{k \neq i} \Gamma_{ik} \\ \nonumber
\mathbb{L}^{\Gamma}_{jj;ii} &=& \Gamma_{ij} \\ \nonumber
\mathbb{L}^{\Gamma}_{ij;ij} &=& -\frac{1}{2}(\sum_{k \neq i} \Gamma_{ik} + \sum_{k \neq j} \Gamma_{jk} )
\end{eqnarray}
with all other elements zero.  

In what follows, it will be convenient to partition the matrix $\mathbb{L}^{\Gamma}$ into its positive
and negative pieces, so that
\begin{equation}
\label{eq:split}
\mathbb{L}^{\Gamma}= \mathbb{L}^{+\Gamma} + \mathbb{L}^{-\Gamma}
\end{equation}
with $\mathbb{L}^{+\Gamma}$ consisting of the terms specified by the second line of
eq. \ref{eq:def_L} and $\mathbb{L}^{-\Gamma}$ comprised of the remaining terms from
the first and third lines.

One final important point is that while eq. \ref{eq:with_q_field_rem} provides effective dynamics
for the system with implications of field fluctuations handled implicitly, this dynamics will still be
far too complicated for exact practical treatment when the system of interest is composed of a 
chromophore embedded in a condensed phase.  The problem is simply one of a complex dynamics
associated with a quantum mechanical many body system.  When it is possible to make some 
effective separation between the relevant part of the system and a weakly coupled (and fast)
bath this problem can be overcome in exactly the same method employed to remove the radiative
field from explicit consideration.  Writing
\begin{equation}
\label{eq:sys_ham}
\hat{H}^{sys} = \hat{H}^{ch} + \hat{H}^{b} + \hat{V}
\end{equation}
for a ``system'' Hamiltonian composed of two parts: $ch$ (the chromophore which 
is directly coupled to the applied field) and $b$ (the bath) weakly coupled
by $\hat{V}$,
we arrive at an equation of motion for the reduced chromophore density matrix
$\sigma$ through application of standard Redfield theory \cite{blum,schatz}
\begin{equation}
\label{eq:full_evolution}
\dot{\sigma}_{ij}(t) = -i\omega_{ij}\sigma_{ij} +\mathbb{L}^{E}_{ij;kl}(t)\sigma_{kl} +\mathbb{L}^{\Gamma}_{ij;kl}\sigma_{kl} +
\mathbb{R}_{ij;kl}\sigma_{kl} \equiv \mathbb{L}_{ij;kl}(t)\sigma_{kl}.
\end{equation}
Here, the matrix $\mathbb{R}$ is the usual Redfield matrix to account for bath perturbations
on the chromophore and the matrix $\mathbb{L}(t)$ reflects the entire dynamics for $\sigma$.  
We note that additivity of contributions stemming from quantum field,
bath and classical (laser) field perturbations to the dynamics of the chromophore should be
viewed as an approximation of ``independent rates of variation" \cite{ct}.  We 
neglect frequency shifts of the chromophore due to $\hat{V}$, so that the labels $ij$ now
correspond to eigenstates of $H^{ch}$.  We consider this
set of approximations as the natural extension of the optical Bloch equations to multi-level
systems in a condensed phase.  Specification of the matrices $\mathbb{L}^E$,  $\mathbb{L}^{\Gamma}$
and $\mathbb{R}$ will allow us to apply this formalism to various physical problems and several
model systems will be considered in the following sections.

\subsection{Extraction of photon counting moments}
\label{sec:extract}
Extending the picture of the preceding section to calculation of photon counting
statistics for single molecule measurements proceeds in a manner analogous
to the case for two level chromophores \cite{cook,zheng2}.  The formal solution
has been presented in ref. \cite{muk1} and we present here a brief derivation following
ref. \cite{zheng2} to clarify our notation and to extend this picture to the calculation of
photon counting moments for individual spontaneous emission transitions (as will
be useful in spectrally resolved emission spectroscopy).

Imagine a detector capable of differentiating between photons that are emitted
for particular chromophore transitions.  In certain cases this would be possible by
only selecting photons within a certain frequency window, in other cases this might not
be experimentally feasible but should be regarded as a gedanken experiment.   
That portion of $\mathbb{L}(t)$ responsible for
placing the chromophore in a lower energy state immediately following the transition
of interest is of special importance for calculating statistics associated  with this transition.  
From eq. \ref{eq:def_L} this is the element $\mathbb{L}^{+\Gamma}_{bb;aa}$
with the numerical value $\Gamma_{ab}$, assuming that we are following 
$a \rightarrow b$ emissions.  Partition eq. \ref{eq:full_evolution} 
to give this single part of the evolution a unique status
\begin{equation}
\dot{\sigma}_{ij} (t) = \mathbb{L}_{ij;kl}'(t)\sigma_{kl} + \Gamma_{ab}\delta_{ij,bb}\delta_{kl;aa}\sigma_{kl}
\equiv \mathbb{L}_{ij;kl}'(t)\sigma_{kl} + \mathbb{L}^{+\Gamma_{ab}}_{ij;kl}\sigma_{kl}
\end{equation}
where $\mathbb{L}'(t)$ is that portion of $\mathbb{L}(t)$ not pulled out in $\mathbb{L}^{+\Gamma_{ab}}$.
In exact analogy to the case with only a two level chromophore, it is the operator 
$\mathbb{L}^{+\Gamma_{ab}}$ that dictates when an $a \rightarrow b$ spontaneous emission event occurs.
Following exactly the same arguments as in ref. \cite{zheng2} allows us to write
\begin{equation}
\label{eq:density_n}
\dot{\sigma}_{ij}^{(n)}(t)=\mathbb{L}_{ij;kl}'(t)\sigma_{kl}^{(n)}+\mathbb{L}^{+\Gamma_{ab}}_{ij;kl}\sigma_{kl}^{(n-1)}
\end{equation}
where $\sigma^{(n)}$ is that portion of the reduced density matrix corresponding to systems that
have previously emitted exactly $n$ photons via $a\rightarrow b$ transitions.

To facilitate the extraction of photon counting moments, we introduce a generating function
version of eq. \ref{eq:density_n}
\begin{eqnarray}
\label{eq:gen_eom}
\dot{\mathcal{G}}_{ij}(t,s) &=& \mathbb{L}_{ij;kl}'(t)\mathcal{G}_{kl}(t,s) + s\mathbb{L}^{+\Gamma_{ab}}_{ij;kl}
\mathcal{G}_{kl}(t,s) = \mathbb{L}_{ij;kl}(t,s)\mathcal{G}_{kl}(t,s) \\ \nonumber
\mathcal{G}(t,s) &\equiv& \sum_{n=0}^{\infty} s^n \sigma^{(n)} (t). 
\end{eqnarray}
The actual generating function for $a \rightarrow b$ photon emissions is obtained by summing
over all ``population'' elements of $\mathcal{G}(s,t)$
\begin{equation}
G(s,t) = \sum_{i}\mathcal{G}_{ii}(s,t)
\end{equation}
which allows for the usual extraction of probabilities for $n$ photon emissions \cite{vankampen}
\begin{equation}
p_n (t) = \frac{1}{n!} \left . \frac{\partial^n}{\partial s^n} G(s,t)\right |_{s=0}
\end{equation}
and factorial moments \cite{vankampen}
\begin{equation}
\label{eq:fac_mom}
\langle n^{(m)} \rangle (t) \equiv \langle n(n-1)(n-2)\ldots (n-m+1)\rangle (t)=\left . \frac{\partial^m}{\partial s^m} G(s,t) \right |_{s=1}.
\end{equation}
 
 Our primary concern in this work shall be the calculation of moments.  To this end, we differentiate
 eq. \ref{eq:gen_eom} with respect to $s$ yielding equations for the $\partial^{m} /\partial s^m
 \mathcal{G}$ elements which, when summed over population elements, yield the moments 
 (when $s=1$).  
 \begin{equation}
 \label{eq:get_moments}
 \frac{\partial }{\partial t} \left ( \frac{\partial^m \mathcal{G}(s,t)}{\partial s^m} \right )=
\mathbb{L}(t,s) \left (\frac{\partial^m \mathcal{G}}{\partial s^m} \right )+ m \mathbb{L}^{+\Gamma_{ab}}
\left (\frac{\partial^{m-1} \mathcal{G}}{\partial s^{m-1}} \right )
\end{equation}
The high order derivatives are dependent upon all lower derivatives as can be seen by iterating
this equation.  For example, moments up to and including second order are generated by solving 
the set of equations
\begin{equation}
\label{eq:comp_mom}
\left( \begin{array}{c}
	\dot{\mathcal{G}}(s,t) \\
	\frac{\partial \dot{\mathcal{G}}(s,t)}{\partial s} \\
	\frac{\partial^2 \dot{\mathcal{G}}(s,t)}{\partial s^2}
	\end{array} \right ) =
	\left ( \begin{array}{ccc}
		\mathbb{L}(t,s) & 0 & 0 \\
		\mathbb{L}^{+\Gamma_{ab}}(s) & \mathbb{L}(t,s) & 0 \\
		0 & 2\mathbb{L}^{+\Gamma_{ab}}(s) & \mathbb{L}(t,s)
		\end{array} \right ) \cdot
	\left ( \begin{array}{c}
		\mathcal{G}(s,t) \\
	\frac{\partial \mathcal{G}(s,t)}{\partial s} \\
	\frac{\partial^2 \mathcal{G}(s,t)}{\partial s^2} \end{array} \right ).
\end{equation}
Evaluation at $s=1$ provides the moments up to second order by way of eq. \ref{eq:fac_mom}.
Since $\mathbb{L}(t,s)$ and $\mathbb{L}^{+\Gamma_{ab}}$ are $N^2 \times N^2$ matrices for a quantum system with $N$ states, the above expression corresponds to solving $3N^2$ coupled equations.
In the cases considered in this work, we will take $E(t)$ to have sinusoidal time dependence so that
the explicit time dependence within $\mathbb{L}(t)$ may be removed by moving to a rotating reference 
frame and applying the rotating wave approximation (RWA).  In this case, solution of these equations
is easily accomplished by directly exponentiating the $3N^2 \times 3N^2$ matrix as outlined
in the next section.  
Equation \ref{eq:comp_mom} is central to all results in this paper and, in principle, could have been
directly solved to reproduce all the calculations presented below.  In practice, we used a numerically
simpler scheme to obtain our results derived from eq. \ref{eq:comp_mom}.  This numerical technique
is elaborated on in sec. \ref{sec:numeric}.  Formation of the matrices $\mathbb{L}(t,s)$ and
$\mathbb{L}^{+\Gamma_{ab}}$ for use in any numerical scheme follow from the preceding section.  
Specific choices for these matrices depend upon the physical systems
under consideration and will be detailed with presentation of our chosen applications.  

The above derivation has assumed that we are interested in the statistics of photons emitted
from one particular chromophore transition ($a \rightarrow b$).  When we are interested in broadband
detection with all photons counted equivalently, the structure of eq. \ref{eq:comp_mom} remains unchanged.
However, the matrices $\mathbb{L}(t,s)$ and $\mathbb{L}^{+\Gamma_{ab}}$ have different forms.  
In that case we substitute $\mathbb{L}^{+\Gamma}$ for $\mathbb{L}^{+\Gamma_{ab}}$ and $\mathbb{L}(t,s)$
is now the matrix formed by appending $s$ to every spontaneous emission matrix element within
$\mathbb{L}(t)$ having a positive sign (i.e. the whole of $\mathbb{L}^{+\Gamma}$).  Calculation of moments for photons associated with some
subset of transitions
(perhaps transitions inside a certain frequency window) proceeds by generalizing to
placement of $s$ variables only on the elements associated with the relevant transitions and making
the corresponding changes to $\mathbb{L}^{+\Gamma}$.  In principle, we could introduce 
a number of different auxiliary variables - each variable corresponding to a particular transition
or subset of transitions.  This leads to expressions for
cross correlations between various transitions.  The extension is straightforward, 
but not explicitly presented here as we do not calculate any such cross correlations in this work.

\subsection{Model Hamiltonians and practical considerations}
\label{sec:practical}
In this work we shall be concerned exclusively with model systems consisting
of a chromophore with two electronic states (ground $|g\rangle$ and excited $|e\rangle$),
so that 
 \begin{equation}
\label{eq:systemh}
\hat{H}^{ch}=|g\rangle H_g \langle g| + |e \rangle H_e \langle e|.
\end{equation}
$H_g$ and $H_e$ are, respectively, the chromophore Hamiltonians for
nuclear motion within the ground and excited states, with eigenfunctions and eigenvalues
specified by
\begin{eqnarray}
\label{hegeigen}
H_g |n_g \rangle & = & \epsilon_{n_g} |n_g \rangle, \\\nonumber
H_e |m_e \rangle & = & \epsilon_{m_e} |m_e \rangle,
\end{eqnarray}
for $m_e=1\ldots N_e$, $n_g=1,\ldots N_g$. In our numerical applications, we consider only a 
finite number of eigenstates associated with nuclear motion, and adopt the 
convention here. 
The nuclear ground  state in the excited manifold is assumed to lie higher
in energy than the nuclear ground state of the ground manifold by an amount
$\hbar \omega_{eg}$.  It is
to be understood that this chromophore Hamiltonian dictates dynamics in
the sense implied by eq. \ref{eq:sys_ham}.  $\hat{H}^{ch}$ is responsible for
the evolution that we designate to be the most important to chromophore dynamics.
The effect of the environment (bath) will be felt through coupling dictated by $\hat{V}$.

Interactions with the radiation field depend upon the matrix elements of
the system's dipole moment operator as evidenced by eq. \ref{eq:emit_rate}
and the presence of $\hat{\boldsymbol{\mu}}$ in $\mathbb{L}^{E}(t)$.  We treat these matrix
elements in the Condon approximation \cite{mukbook} such that
\begin{equation}
\label{eq:condon}
D_{n_g;m_e} =\langle g| \hat{\boldsymbol{\mu}}|e \rangle \langle n_g|m_e \rangle \equiv \boldsymbol{\mu}_{0}\langle n_g|m_e \rangle.
\end{equation}
The dipole operator is assumed to act solely in the electronic space with only
off-diagonal coupling between ground and excited states.  Individual transition intensities are mediated by the
overlap of nuclear wavefunctions.
We always consider a monochromatic exciting field of constant intensity and polarization direction, 
so that
\begin{equation}
\mathbf{E}(t)=\boldsymbol{\mathcal{E}}_0 \cos (\omega_L t).
\end{equation}
For future notational simplicity we define constants $\Gamma_0$ and
$\Omega_0$ as
\begin{eqnarray}
\Gamma_0 &=& \frac{\omega_{eg}^3 |\boldsymbol{\mu}_{0}|^2}{3\pi \epsilon \hbar c^3} \\ \nonumber
\Omega_0 &=& \boldsymbol{\mathcal{E}}_0\cdot\boldsymbol{\mu}_{0}/\hbar 
\end{eqnarray}
These constants represent the spontaneous emission rate and Rabi frequency for an electronic transition
between states with perfect overlap of nuclear wavefunctions.  

These definitions allow us specify the form
of matrices $\mathbb{L}^{E}(t)$ and $\mathbb{L}^{\Gamma}$.  $\mathbb{L}^{\Gamma}$ follows immediately from
 eq. \ref{eq:def_L}.  All we need are the emission rates $\Gamma_{ij}$ for
all $i \rightarrow j$ transitions.  Since our models only allow transitions between excited and ground electronic
states we need only consider rates of the form $\Gamma_{\ket{e}\ket{m_e};\ket{g}\ket{n_g}} \equiv \Gamma_{m_en_g}$ with values
\begin{equation}
\label{eq:gam_mn}
\Gamma_{m_en_g}=\Gamma_0 \left | \bra{m_e}n_g\rangle \right |^2.
\end{equation}
All positions in $\mathbb{L}^{\Gamma}$ diagonal in the electronic subspace are necessarily zero due to our assumptions
about the dipole operator, so the above completely specifies the $\mathbb{L}^{\Gamma}$ matrix.

Formation of $\mathbb{L}^E (t)$ is slightly more complicated due to the nature of the coupling to the applied field, which
makes for a matrix less sparse than the emission matrix.  We first realize that, as in the usual optical Bloch equations,
density matrix elements diagonal in the electronic subspace are coupled to those off-diagonal in the electronic
subspace and vice versa.  Also, by analogy to the optical Bloch equations we retain only those terms corresponding
to resonant excitation by the field (i.e. a photon is absorbed and electronic state rises or a photon is emitted and 
state drops) by invoking the Rotating wave approximation (RWA)  \cite{ct}.  We make use of the definition
\begin{equation}
\label{eq:omg_mn}
\Omega_{m_e n_g}=\Omega_0 |\bra{m_e} n_g\rangle|
\end{equation}
to give the elements of $\mathbb{L}^E (t)$ within the RWA
\begin{eqnarray}
\label{eq:driving}
\mathbb{L}^E_{n_gm_g;k_el_g} &=&  -\overline{\mathbb{L}^{E}_{k_el_g;n_gm_g}} = +\frac{i}{2}\Omega_{n_gk_e}e^{i\omega_L t}  \delta_{m_g,l_g}\\ \nonumber
\mathbb{L}^E_{n_gm_g;k_gl_e} &=& -\overline{\mathbb{L}^E_{k_gl_e;n_gm_g}} = -\frac{i}{2}\Omega_{ l_em_g}e^{-i\omega_L t} \delta_{n_g,k_g}\\ \nonumber
\mathbb{L}^E_{n_em_e;k_el_g} &=&-\overline{\mathbb{L}^E_{k_el_g;n_em_e}}=-\frac{i}{2}\Omega_{l_gm_e}e^{i\omega_L t} \delta_{n_e,k_e}\\ \nonumber
\mathbb{L}^E_{n_em_e;k_gl_e} &=& -\overline{\mathbb{L}^E_{k_gl_e;n_em_e}}=+\frac{i}{2}\Omega_{n_ek_g}e^{-i\omega_L t} \delta_{m_e,l_e}. \\ 
\end{eqnarray}
Over bars represent complex conjugation.  (The above definitions assume that our dipole operator
matrix elements are real quantities.  In the presence of a magnetic field this condition could be violated,
but we restrict attention away from such cases.)

The only portion of $\mathbb{L}(t)$ remaining to be specified is the Redfield matrix
for transitions of the chromophore induced by environmental bath fluctuations,
$\mathbb{R}$.  The relaxation matrix is given by \cite{blum,schatz}
\begin{equation}
\label{eq:relaxm}
\mathbb{R}_{i j; k l}  =  -\delta_{i k} \sum_r t_{l r r j}^{-} - 
                    \delta_{l j} \sum_r t_{i r r k}^{+} +
                t_{j l i k}^{-} +  
                t_{j l i k}^{+},
\end{equation}
where
\begin{eqnarray}
\label{eq:t12}
t_{l j i k}^{+} & = & \frac{1}{\hbar^2} \int_{0}^{\infty} d \tau e^{-i \omega_{i k} \tau}
                 \langle \hat{V}_{ l j}(\tau) \hat{V}_{i k}(0) \rangle_{b}, \\\nonumber
t_{l j i k}^{-} & = & \frac{1}{\hbar^2} \int_{0}^{\infty} d \tau e^{-i \omega_{l j} \tau}
                 \langle \hat{V}_{ l j}(0) \hat{V}_{i k}(\tau) \rangle_{b}, 
\end{eqnarray}
are Fourier-Laplace transforms of the correlation functions of the system and bath coupling at the specified frequency. The bath-space
Heisenberg operators are defined by
\begin{eqnarray}
\label{eq:hsbmatrix}
\hat{V}_{ k i}(\tau) & = & e^{\frac{i}{\hbar} \hat{H}^b \tau} \hat{V}_{ k i} e^{-\frac{i}{\hbar} \hat{H}^b \tau}, \\\nonumber
 \hat{V}_{ k i} & = &  \langle k | \hat{V} | i \rangle
\end{eqnarray} 
and the averages $\langle \ldots \rangle_b$ specify a thermal average over bath degrees of freedom only.
In all models we consider, bath fluctuations are capable of causing transitions between levels within a 
particular electronic state, but are not permitted to induce radiationless transitions between electronic
states.  Further discussion on the evaluation of $\mathbb{R}$ will appear
in sections \ref{sec:tls} and \ref{sec:harmonic} as specific models for chromophore and
bath are introduced.

Given the particular form of our model systems, it is highly beneficial to solve eq. \ref{eq:full_evolution}
in a rotating reference frame by introducing new variables
\begin{eqnarray}
\tilde{\sigma}_{n_gm_e} &=&\sigma_{n_gm_e}e^{-i\omega_L t} \\ \nonumber
\tilde{\sigma}_{n_em_g} &=&\sigma_{n_em_g}e^{i\omega_L t} \\ \nonumber
\tilde{\sigma}_{n_em_e} &=& \sigma_{n_em_e} \\ \nonumber
\tilde{\sigma}_{n_gm_g} &=& \sigma_{n_gm_g}. 
\end{eqnarray}
The primary advantage of this formulation being that eq. \ref{eq:full_evolution} is
recast in a form without explicit time dependence
\begin{equation}
\label{eq:full_rotate}
\dot{\tilde{\sigma}}_{ij}(t) = -i\mathbb{W}_{ij;kl}\tilde{\sigma}_{kl} +\mathbb{L}^{E}_{ij;kl}\tilde{\sigma}_{kl} +\mathbb{L}^{\Gamma}_{ij;kl}\tilde{\sigma}_{kl} + \mathbb{R}_{ij;kl}\tilde{\sigma}_{kl} \equiv \mathbb{L}_{ij;kl} \tilde{\sigma}_{kl}
\end{equation}
where the diagonal matrix $\mathbb{W}$ is given by
\begin{eqnarray}
\mathbb{W}_{n_gm_g;n_gm_g} &= &\omega_{n_gm_g} \\ \nonumber
\mathbb{W}_{n_em_e;n_em_e} &=& \omega_{n_em_e} \\ \nonumber
\mathbb{W}_{n_em_g;n_em_g} &=& \omega_{n_em_g} - \omega_L \\ \nonumber
\mathbb{W}_{n_gm_e;n_gm_e} &=& \omega_{n_gm_e} + \omega_L. 
\end{eqnarray}
The matrix $\mathbb{L}^E$ is simply the matrix specified by eq. \ref{eq:driving}, evaluated at $t=0$ and
the remaining matrices are unchanged relative to the original basis.  Since the populations of $\tilde{\sigma}$ are identical to
$\sigma$, we may calculate photon emission statistics using these transformed variables without
any changes to the formalism of the preceding subsection.  In particular, we may calculate
eq. \ref{eq:comp_mom} as
\begin{equation}
\label{eq:comp_mom_rotate}
\left( \begin{array}{c}
	\dot{\tilde{\mathcal{G}}}(s,t) \\
	\frac{\partial \dot{\tilde{\mathcal{G}}}(s,t)}{\partial s} \\
	\frac{\partial^2 \dot{\tilde{\mathcal{G}}}(s,t)}{\partial s^2}
	\end{array} \right ) =
	\left ( \begin{array}{ccc}
		\mathbb{L}(s) & 0 & 0 \\
		\mathbb{L}^{+\Gamma_{ab}}(s) & \mathbb{L}(s) & 0 \\
		0 & 2\mathbb{L}^{+\Gamma_{ab}}(s) & \mathbb{L}(s)
		\end{array} \right ) \cdot
	\left ( \begin{array}{c}
		\tilde{\mathcal{G}}(s,t) \\
	\frac{\partial \tilde{\mathcal{G}}(s,t)}{\partial s} \\
	\frac{\partial^2 \tilde{\mathcal{G}}(s,t)}{\partial s^2} \end{array} \right ).
\end{equation}
where the time independent $\mathbb{L}$ is specified by eq. \ref{eq:full_rotate}
and $\tilde{\mathcal{G}}(s,t)$ is given by
\begin{eqnarray}
\label{eq:gen_eom_rotate}
\tilde{\mathcal{G}}(t,s) &\equiv& \sum_{n=0}^{\infty} s^n \tilde{\sigma}^{(n)} (t). 
\end{eqnarray}
Summing over the ``population'' elements of $\tilde{\mathcal{G}}$ still 
returns the original generating function for photon emissions, $G(s,t)$,
so calculations in this frame return emission statistics equivalent to
the original formulation.  Numerics in this basis are preferred, since
eq. \ref{eq:comp_mom_rotate} may be solved simply by direct matrix
exponentiation.

\subsection{Reported quantities and numerical details}
\label{sec:numeric}
The bulk of the preceding sections has been devoted to establishing
models for reduced chromophore dynamics, i.e. how to specify the superoperator
matrix $\mathbb{L}(t)$ in eq. \ref{eq:full_evolution} or the corresponding time-independent
matrix $\mathbb{L}$ in eq. \ref{eq:full_rotate}.  Given this matrix, it is a trivial programming
task to extend the standard calculation of density matrix evolution to photon counting 
observables.  The matrix $\mathbb{L}(s)$ is formed by appending the auxiliary variable $s$
to elements of $\mathbb{L}^{+\Gamma}$  reflecting spontaneous
emission transitions of interest.  In the case of a single relevant transition, only one element
is modified.  In broadband detection we append an
$s$ to the entire $\mathbb{L}^{+\Gamma}$ matrix.  Given $\mathbb{L}(s)$, the block form of eq. \ref{eq:comp_mom_rotate}
follows immediately and calculation of $\tilde{\mathcal{G}}$ is provided by simple matrix
exponentiation.  Summing over population elements of $\partial^m \tilde{\mathcal{G}}(s,t)/\partial s^m$ for $s=1$ yields 
the factorial photon counting moment of order $m$.  Although the matrix in \ref{eq:comp_mom_rotate} is specific
to calculation of $m=2$, higher order moments can be calculated in analogous
fashion by extending the block matrix as implied by eq. \ref{eq:comp_mom}. Since we assume no photon emissions prior
to $t=0$, the initial condition employed in eq. \ref{eq:comp_mom_rotate} is simply
$\tilde{\mathcal{G}}_{ij}(s,0)=\tilde{\sigma}_{ij}(0)$ with all $s$ derivatives of $\tilde{\mathcal{G}}$ equal
to zero.

The moments reported in this work will be presented in terms of absorption and emission lineshapes
and corresponding Mandel  $Q$ parameter \cite{optlett} spectra.  Mandel's $Q$ parameter is related
to the factorial moments via
\begin{equation}
\label{eq:manq}
Q(t) \equiv \frac{ \langle n^2 \rangle(t) - \langle n \rangle^2(t)}{\langle n \rangle(t)}-1, 
\end{equation}
and serves as a convenient means to report second order photon statistics.  Positive $Q$ values
reflect photon bunching behavior (an elevated variance in $n$ relative to Poisson processes with the same mean), 
negative $Q$ values anti-bunching behavior (diminished variance in $n$ relative to a Poisson process with
the same mean) and $Q=0$ is consistent with purely Poissonian statistics.

Energy conservation implies that we may calculate absorption lineshapes, by counting the relative rate of
photon emission (photons from all transitions are counted)  as a function of the exciting frequency
\begin{equation}
\label{eq:lineshape}
I(\omega_L)=\lim_{t\rightarrow \infty} \frac{\partial}{\partial t}\langle n \rangle (t) 
\equiv \lim_{t\rightarrow \infty} \frac{\partial}{\partial t}
    \left[ \left. \frac{\partial}{\partial s} G(s,t) \right|_{s=1} \right].
\end{equation}
Every emitted photon corresponds to a prior excitation of the chromophore, and hence a
quantum of energy ($\hbar \omega_L$) extracted from the incident field.  
We evaluate lineshapes in the limit of long times to insure
that the system is in a steady state.  The time dependence of $d\langle n \rangle/dt$ at early times
is interesting as well \cite{zheng1,zheng2}, but not specifically considered in this work.  The $Q$ parameter absorption spectra are calculated  in analogous fashion, although the definition
of $Q$, with $\langle n \rangle (t)$ in the denominator, insures saturation to a constant value as time
becomes large.  It is unnecessary to take a time derivative to report a meaningful quantity here and
the $Q$ parameter itself as a function of exciting frequency is reported.  Again, in the ``absorption
$Q$ spectra" we collect all photon emissions (broadband detection).

Emission lineshapes and $Q$ parameter are calculated in similar fashion, but we resolve the photon
statistics by frequency of the emitted photons.  More precisely, we resolve by the transitions
the photons originate from.  In the cases we consider, the allowed transitions are either well resolved
in frequency (frequency differences much larger than natural linewidths) or perfectly degenerate, so that 
there is no ambiguity in assigning photons to a particular frequency ``window''.  We report our results
as
\begin{equation}
\label{eq:lineshape1}
I(\omega_{E}; \omega_L)=\lim_{t\rightarrow \infty} \frac{\partial}{\partial t}\langle n_{\omega_{ij}=\omega_E} \rangle (t) 
\end{equation}
The above notation specifies that we only consider photons from transitions on resonance with $\omega_E$.  Collection
of these statistics follows the prescription previously described.  The matrix $\mathbb{L}(s)$ depends on 
$\omega_E$ as placement of $s$ variables is dictated by which transitions are on resonance with $\omega_E$.
We note that our emission ``spectra'' are thus not quite spectra in the usual sense.  Our spectral lines are
infinitely sharp, without broadening (see fig. \ref{fig:tlsemi}).  In principle, we could artificially broaden these lines
by making them Lorentzians with the natural linewidth of each transition, but we have not done so.  What our calculations
directly provide are the statistics associated with particular molecular transitions, not the actual frequency of the emitted
photons.  Note that our lineshapes will also, in general, depend upon the frequency of the exciting light as different
excitations can lead to different steady state populations of the chromophore.

The $Q$ parameter emission spectrum follows similarly
\begin{equation}
\label{eq:manqi}
Q(\omega_E; \omega_L) \equiv \frac{ \langle n^2_{\omega_{ij}=\omega_E} \rangle - 
\langle n_{\omega_{ij}=\omega_E} \rangle^2}{\langle n_{\omega_{ij}=\omega_E} \rangle}-1 
\end{equation}
where we stress that the photon numbers $n$ collected above reflect only those photons
stemming from transitions on resonance with $\omega_E$.

For multi level quantum systems the matrix of eq. \ref{eq:comp_mom_rotate} can become very large ($3N^2\times 3N^2$ for $N$ quantum levels).  If moments higher than second order are desired,
the matrix becomes even bigger.  Direct exponentiation of such matrices over
a wide range of frequencies is computationally expensive and, for sufficiently large N and/or
moment order, eventually
becomes computationally intractable. In this work we focus on statistics calculated in 
the long time (steady state) limit.  For direct exponentiation, this limit has the 
additional computational complications associated with the identification of a time sufficiently 
large for the steady state to be attained, yet sufficiently small to insure numerical stability.  
When only steady state information is desired, analytical progress can be made on eq. \ref{eq:comp_mom_rotate}, allowing calculation to proceed via diagonalization of matrices no larger than $N^2\times N^2$ and without the need to identify a suitable finite time at which the long time limit is
reached.  The calculation is summarized below.

The equations of motion for $\tilde{\mathcal{G}}$ and its
$s$ derivatives (Eq. \ref{eq:comp_mom_rotate}) can be formally
integrated to yield
\begin{eqnarray}
\label{eq:integrated}
\tilde{\mathcal{G}}'(t) &=&   \int_{0}^{t}dt' e^{\mathbb{L}(t-t')}
\mathbb{L}^{+\Gamma} \rho_{s.s.} \\ \nonumber
\tilde{\mathcal{G}}''(t) &=&  2\int_{0}^{t}dt' e^{\mathbb{L}(t-t')}
\mathbb{L}^{+\Gamma}\int_{0}^{t'}dt'' e^{\mathbb{L}(t'-t'')} \mathbb{L}^{+\Gamma} \rho_{s.s}.
\end{eqnarray}
Here we have assumed that the system began in the steady state at $t=0$  and that we began
counting photons at $t=0$ (different initial conditions lead to negligible 
corrections in the long time limit).  We have introduced
a prime notation for $s$ derivatives (i.e. $\frac{\partial \tilde{\mathcal{G}}}{\partial s} \equiv 
\tilde{\mathcal{G}}'$) and we have evaluated everything for $s=1$.  The steady state limit for
the density matrix $\rho_{s.s.}$ is expected on physical grounds for systems driven by 
external perturbations and allowed to relax via radiative and non-radiative transitions - its
existence was verified for the model systems studied in this work.

The matrix $\mathbb{L}$ may be diagonalized and we write $\Lambda = \chi^{-1}\mathbb{L}\chi$
with $\Lambda$ the diagonal representation of $\mathbb{L}$.  The columns of $\chi$ consist
of the right eigenvectors of $\mathbb{L}$ and the rows of $\chi^{-1}$ are the left eigenvectors
of $\mathbb{L}$.  The associated eigenvalues of $\mathbb{L}$ are complex numbers with negative real
parts, excepting the single eigenvalue associated with the steady state which is zero.  Ordering the 
eigenvalues $\{ \lambda_{s.s} =0, \lambda_2, \lambda_3,\ldots \}$, so that
\begin{equation}
\Lambda =
	\left( \begin{array}{cccc}
         0 & 0 & 0 &\cdots \\
         0 & \lambda_2 & 0 &\cdots \\
         0 & 0 & \lambda_3 &\cdots \\
         \vdots & \vdots & \vdots &\ddots
           \end{array}  \right)
\end{equation}
we see that it is possible to partition the time evolution operator 
$U(\tau)=e^{\mathbb{L}\tau}\equiv U_0 + U_1(\tau)$
into two pieces such that the first corresponds to the (lack of) evolution of the steady state
and the second piece reflects all other dynamics in the system.
\begin{displaymath}
U_0= 
\chi\left( \begin{array}{cccc}
         1 & 0 & 0 &\cdots \\
         0 & 0 & 0 &\cdots \\
         0 & 0 & 0 &\cdots \\
         \vdots & \vdots & \vdots &\ddots
           \end{array}  \right)\chi^{-1}, \\
\end{displaymath}
\begin{displaymath}
U_1(\tau)= 
\chi\left( \begin{array}{cccc}
         0  & 0 & 0 &\cdots \\
         0 & e^{\lambda_{2}\tau} & 0 &\cdots \\
         0 & 0 & e^{\lambda_{3}\tau} &\cdots \\
         \vdots &  \vdots & \vdots & \ddots
           \end{array}  \right)\chi^{-1}. \\
\end{displaymath}

Partitioning the matrices in this way allows us to explicitly carry out the integrations in eq. \ref{eq:integrated}
to give (large time limit assumed)
\begin{eqnarray}
\label{eq:longtime}
\tilde{\mathcal{G}}' &=& (tU_{0}+X)\mathbb{L}^{+\Gamma}\rho_{s.s.} \\ \nonumber
\tilde{\mathcal{G}}''  &= & t^{2} (U_{0}\mathbb{L}^{+\Gamma})^{2}\rho_{s.s}+2t(U_{0}\mathbb{L}^{+\Gamma}X\mathbb{L}^{+\Gamma}+X\mathbb{L}^{+\Gamma}U_{0}\mathbb{L}^{+\Gamma})\rho_{s.s.} \\ \nonumber
X &\equiv& \chi\left( \begin{array}{cccc}
         0  & 0 & 0 &\cdots \\
         0 & -\lambda_2^{-1} & 0 &\cdots \\
         0 & 0 & -\lambda_3^{-1} &\cdots \\
         \vdots &  \vdots & \vdots & \ddots
           \end{array}  \right)\chi^{-1}
\end{eqnarray}
The long time (steady state) limit for the rate of photon emission (intensity) and the $Q$ parameter follow immediately 
\begin{eqnarray}
\label{eq:newmethod}
\frac{d\langle n \rangle}{dt} &=&\sum_{P.E.} U_{0}\mathbb{L}^{+\Gamma}\rho_{s.s.} \\ \nonumber
Q &=& 2\frac{\sum_{P.E.} U_{0}\mathbb{L}^{+\Gamma}X\mathbb{L}^{+\Gamma}\rho_{s.s}}{\sum_{P.E.}
U_{0}\mathbb{L}^{+\Gamma}\rho_{s.s.}},
\end{eqnarray}
where the summations are over the population elements of the resulting vectors.  

Eq. \ref{eq:newmethod}
was used in the computation of all quantities reported in the examples discussed below.  We stress that no 
approximations have been introduced into these equations.  The simplifications we obtain are due to the
fact that we only consider the infinite time limit in eq. \ref{eq:newmethod}.    
The numerical advantages of eq. \ref{eq:newmethod}
relative to direct matrix exponentiation are many fold.  First, it is not necessary to pick a time to evaluate
your expressions and somehow confirm that this time is both large enough to insure the steady state yet
small enough to avoid numerical instabilities.  Eq. 
\ref{eq:newmethod} assumes $t \rightarrow \infty$.
Using this method one only has to find the eigenvalues and eigenvectors of the matrix $\mathbb{L}$ for a given excitation frequency to obtain both the
intensity and the Q parameter.  This matrix is three times smaller in linear dimension than the matrix that must
be exponentiated to solve eq. \ref{eq:comp_mom_rotate}.  If higher moments are required, you still have only
to diagonalize the $\mathbb{L}$ matrix for use in expressions similar to eq. \ref{eq:longtime}.  Finally, while
matrix exponentiation requires that you repeat the entire calculation to obtain statistics for various detection
possibilities (broadband, a single transition counted, several transitions counted, etc.), the present scheme
only requires a single diagonalization for all possible detection schemes.  Different detection possibilities
manifest themselves only through the matrix $\mathbb{L}^{+\Gamma}$ which does not have to be diagonalized.
The pieces of eq. \ref{eq:newmethod} dependent on matrix diagonalization ($X$,$U_0$,$\rho_{s.s.}$) do
not vary with different detection schemes.  This is a significant computational advantage when calculating
emission spectra since the bulk of the calculation need only be performed a single time.

\section{Chromophore coupled to a two level system}
\label{sec:tls}

\subsection{Model description}
As a first example, we consider the case of a chromophore coupled to a two level system (TLS).
The two level system model is of interest both for theoretical reasons (it is arguably the simplest
case of dynamics beyond that of an isolated two level chromophore) and also for its utility in 
describing 
the thermal behavior of low temperature glasses \cite{phillips,anderson}.  The model
is also frequently applied to the spectroscopy of chromophores embedded in low temperature glasses
\cite{silb}.  Although TLS dynamics is often treated as a purely
stochastic perturbation of the chromophore system, we adopt a more precise, quantum mechanical picture
here.  The following description of coupled chromophore-TLS dynamics is quite terse.  
We refer readers to the review by Silbey \cite{silb} for more detail on the Redfield dynamics that we
employ.

The nature of TLS dynamics within the glass is presumably the localized rearrangement of a small cluster of 
atoms \cite{phillips,anderson} corresponding to movement between two distinct energy minima.  The coupling
between TLS and chromophore enters as a different effective splitting between chromophore ground and
excited states depending upon which minima the TLS resides in.  Assuming this coupling is due to strain dipole interactions
between chromophore and TLS we expect the interaction to scale as $1/r^3$ in the distance between TLS and
chromophore centers \cite{silb}.  The basis of TLS ``minima'' states is not expected to be diagonal as tunneling may occur
between minima.  In addition, coupling between the TLS and long wavelength phonons in the glass acts as mechanism
for coupling the TLS-chromophore system to its glassy environment.  Adopting the notation of sec. \ref{sec:practical} The mathematical formulation of this picture is \cite{silb,brownsilbey}
\begin{eqnarray}
\label{eq:heg}
H_g&=&-\frac{\hbar \omega_{eg}}{2}+\left( \frac{A}{2}-\frac{\alpha}{4 r^3} \right) \sigma_z^{TLS} + \frac{J}{2} \sigma_x^{TLS},\\\nonumber
H_e&=&+\frac{\hbar \omega_{eg}}{2}+\left( \frac{A}{2}+\frac{\alpha}{4 r^3} \right) \sigma_z^{TLS} + \frac{J}{2} \sigma_x^{TLS} \\\nonumber
\hat{V}&=&\sum_{q} g_q \left( b_{-q}^{\dagger}+b_{q} \right) \sigma_z^{TLS} \\\nonumber
\hat{H}^b &=& \sum_{q}b_q^{\dagger}b_q \hbar \omega_q.
\end{eqnarray}
Here, $A$ and $J$ are respectively the asymmetry and tunneling matrix element 
for the TLS and $\sigma_z^{TLS}$,$\sigma_x^{TLS}$ are Pauli matrices in the basis
of TLS localized ``minima'' states. $\omega_{eg}$ is the chromophore transition frequency in the absence of
interactions. The index $q$
labels the phonon modes of the system and $b_q^{\dagger}$, $b_q$, $\omega_q$
and $g_q$
are the the creation operator, annihilation operator, frequency and 
TLS strain field coupling constants for the $q$th mode.

\begin{figure}
\epsfig{file=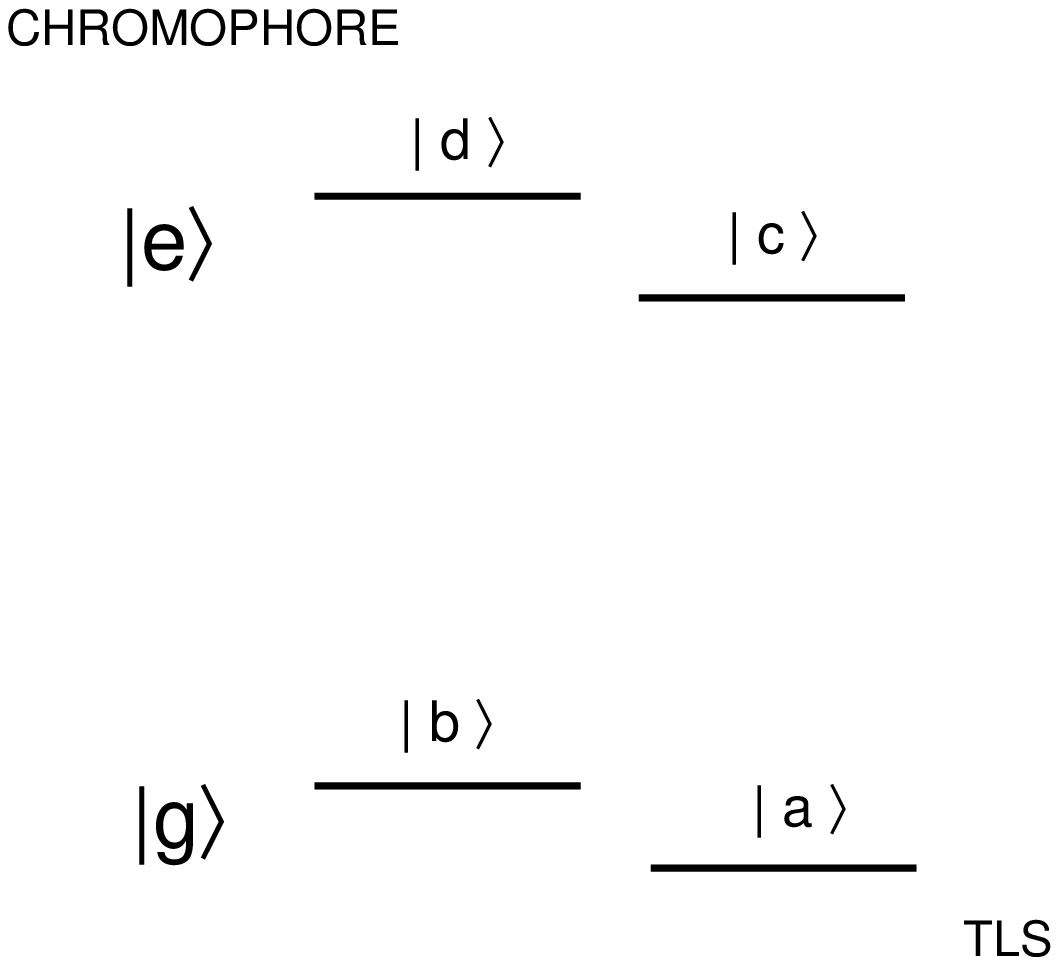, width=3.25in}
\caption{Energy level diagram for the composite chromophore -TLS system}
\label{fig:modeltls}
\end{figure}

We diagonalize the chromophore-TLS portion of our Hamiltonian and label the
four eigenstates  $| a \rangle$, $|b \rangle$ 
$| c \rangle$ and $| d \rangle$ (see fig. \ref{fig:modeltls}) in order of increasing energy (we assume $\hbar \omega_{eg}$
to be by far the largest energy scale in the problem).
In this basis eq.\ref{eq:heg} can be 
written
\begin{eqnarray}
\label{eq:heq1}
H_g&=&\omega_a |a \rangle \langle a| + \omega_b |b \rangle \langle b|, \\\nonumber
H_e&=&\omega_c |c \rangle \langle c| + \omega_d |d \rangle \langle d| \\\nonumber
\hat{V} & = & \sum_q g_q \left( b_{-q}^{\dagger} + b_q \right)
             \left[ \frac{J}{\omega_g} \left( |b \rangle \langle a|
                     + |a \rangle \langle b| \right) \right.+  \\\nonumber
       &    & \left. \frac{J}{\omega_e}  \left( |c \rangle \langle d|
                     + |d \rangle \langle c| \right) \right],
\end{eqnarray}
where $\omega_a$, $\omega_b$, $\omega_c$, $\omega_d$, $\omega_{g}$ and $\omega_{e}$ 
are the frequencies 
\begin{eqnarray}
\label{eq:eigvalue}
\omega_a&=&-\frac{1}{2}\omega_{eg}-\frac{1}{2}\sqrt{J^2+(A-P)^2}, \\\nonumber
\omega_b&=&-\frac{1}{2}\omega_{eg}+\frac{1}{2}\sqrt{J^2+(A-P)^2}, \\\nonumber
\omega_c&=&+\frac{1}{2}\omega_{eg}-\frac{1}{2}\sqrt{J^2+(A+P)^2}, \\\nonumber
\omega_d&=&+\frac{1}{2}\omega_{eg}+\frac{1}{2}\sqrt{J^2+(A+P)^2} \\\nonumber
\omega_g&=& \omega_b - \omega_a \\\nonumber
\omega_e&=&\omega_d - \omega_c
\end{eqnarray}
where we have set $\frac{P}{2}\equiv\frac{\alpha }{4r^3}$.  Note that we have intentionally 
omitted all (system) diagonal contributions to the system-bath coupling since these terms will
yield no contribution to the Redfield matrix.

Specification of $\mathbb{R}$ is quite simple (if tedious) and proceeds by calculating the terms
specified in eqs. \ref{eq:relaxm} and \ref{eq:t12}.  Since the bath is formed by a set of bosons
(phonons), evaluation of the correlation functions is dictated by the well known properties of these operators.
In particular since
\begin{eqnarray}
b_q(t)&=&e^{-i\omega_q t}b_q(0) \\\nonumber
b_q^{\dagger}(t)&=&e^{+i\omega_q t}b_q^{\dagger}(0) \\\nonumber
\langle b_q b_q^{\dagger}\rangle_b &=& (1-e^{-\beta\hbar\omega_q})^{-1} \\\nonumber
\langle b_q^{\dagger}b_q\rangle_b &=& e^{-\beta\hbar\omega_q}(1-e^{-\beta\hbar\omega_q})^{-1}
\end{eqnarray}
the correlation functions become
\begin{equation}
\langle \hat{V}_{ ij}(\tau) \hat{V}_{kl}(0) \rangle_{b}=\sum_q g_q^{ij}g_q^{kl}(1-e^{-\beta\hbar\omega_q})^{-1}
[e^{-i\omega_q\tau}+e^{-\beta\hbar\omega_q}e^{i\omega_q\tau}].
\end{equation}
The coupling constants $g_q$ are chosen to reflect strain field coupling between TLS and the phonon bath
\cite{silb}; they scale with $q$ as $q^{1/2}$.
The $ij$ and $kl$ suffixes on $g_q$ indicate that there are additional constants that need to be included - either
$J/\omega_e$ or $J/\omega_g$ depending upon which specific terms the indices refer to.  Integration in time over
these terms as specified by eq. \ref{eq:t12} serves to create a delta function in frequency which makes evaluation of
the sum over $q$ trivially easy if we approximate the sum as an integral.  By this approach we calculate,
for example,
\begin{eqnarray}
\mathbb{R}_{cc;dd} = e^{\beta\hbar\omega_e}\mathbb{R}_{dd;cc} &=& C\omega_e J^2\frac{1}{1-e^{-\beta\hbar \omega_e}} \\\nonumber
\mathbb{R}_{aa;bb} = e^{\beta\hbar\omega_g}\mathbb{R}_{bb;aa}&=& C\omega_g J^2\frac{1}{1-e^{-\beta\hbar \omega_e}}\\\nonumber
\mathbb{R}_{ca;db} &=& \frac{1}{2}\left [ \frac{\omega_e}{\omega_g}\mathbb{R}_{cc;dd} + \frac{\omega_g}{\omega_e}
\mathbb{R}_{aa;bb} \right ] \\\nonumber
\mathbb{R}_{db;ca} &=& \frac{1}{2}\left [ \frac{\omega_e}{\omega_g}\mathbb{R}_{dd;cc} + \frac{\omega_g}{\omega_e}
\mathbb{R}_{bb;aa} \right ]
\end{eqnarray}
Where $C$ is a collection of constants incorporating the coupling strength between TLS and bath, which is 
typically taken as a parameter used to fit experiment rather than estimated from first principles
\cite{skinner}.  Of course the top two lines just express the phonon assisted transition rates from state $d$ to
$c$ and $b$ to $a$ as expected.
Other elements
follow similarly.  We make no effort to implement the customary secular approximations to these equations
as the equations are solved
numerically and the highly oscillatory terms will remove themselves from consideration naturally.

\subsection{Numerical results}

In this section we present numerical results for the model system described above.  
The framework for calculating the fully quantum dynamical results are spelled out
in sec. \ref{sec:theory}.  Physical constants have been chosen to correspond with
typical situations for a glassy material \cite{skinner,brownsilbey}. In order to
compare with our previous work on stochastic models, it is necessary to map the above 
quantum description to a stochastic picture.  Details for calculating photon statistics for
a stochastic TLS coupled to a chromophore has been presented in detail elsewhere
\cite{zheng2}.  Readers are referred there for a discussion, where we have employed notation
identical to the present work.  Determination of appropriate
model parameters for the stochastic model, based upon the above quantum picture,
is well established \cite{silb}.  In the stochastic picture the TLS acts solely to modulate
the transition frequency of the chromophore, causing hops between $\omega_{eg}+\nu$
and $\omega_{eg} - \nu$.  The rate of hopping is given by $R_{\uparrow}$ for transitions
to the less thermally occupied TLS state and $R_{\downarrow}$ for the reverse direction.
The difference in energy of the two TLS states is provided by detailed balance.
Correspondence with the quantum model is accomplished by
\begin{eqnarray}
\label{eq:nuexp0}
\nu & = & \frac{1}{2} (\omega_e - \omega_g)\\\nonumber
R_{\uparrow} & = &  CEJ^2 \frac{e^{-\beta E}}{1-e^{-\beta E}}  \\\nonumber
R_{\downarrow} & = &  CEJ^2 \frac{1}{1-e^{-\beta E}} \\\nonumber
E & = & \sqrt{A^2+J^2}.
\end{eqnarray}

The idea of the stochastic approach is that coupling between TLS and chromophore
only manifests itself through modulation of the absorption frequency of the chromophore as modulated by TLS hops.
TLS dynamics and thermal properties are completely unaffected by the chromophore, hence the total
independence of TLS energy scale and flip rates on chromophore properties - i.e. these quantities are
calculated by setting the TLS-chromophore coupling constant $\alpha$ to zero in our earlier expressions.
Of course it is crucial to keep $\alpha$ in the frequencies, otherwise the TLS would have no effect on the
chromophore at all.  The stochastic approximation is expected to work quite well when $\alpha$ is small.
in that case transition elements of the Redfield matrix are well approximated by using rates inferred from
eq. \ref{eq:nuexp0}.  It should be noted that the stochastic approach is obviously deficient in one sense.
There are four possible transition frequencies implied by the quantum level  diagram in fig. \ref{fig:modeltls} 
and the stochastic picture only predicts two.  For small $\alpha$ and/or large $r$, half the transitions rarely occur
because of poor Franck-Condon overlap.  Given our notation, the transitions $c\rightarrow a$ and 
$d \rightarrow b$ are the strong ones (assuming weak coupling).  At high couplings strengths, 
half of the transitions will necessarily be missed by the stochastic picture.  The following numerical
examples highlight both the practicality of the present fully quantum approach in calculations as well as
the shortcomings of the popular stochastic approximation over certain parameter regimes.

\subsubsection{Weak coupling between chromophore and TLS.}

\begin{figure}
\epsfig{file=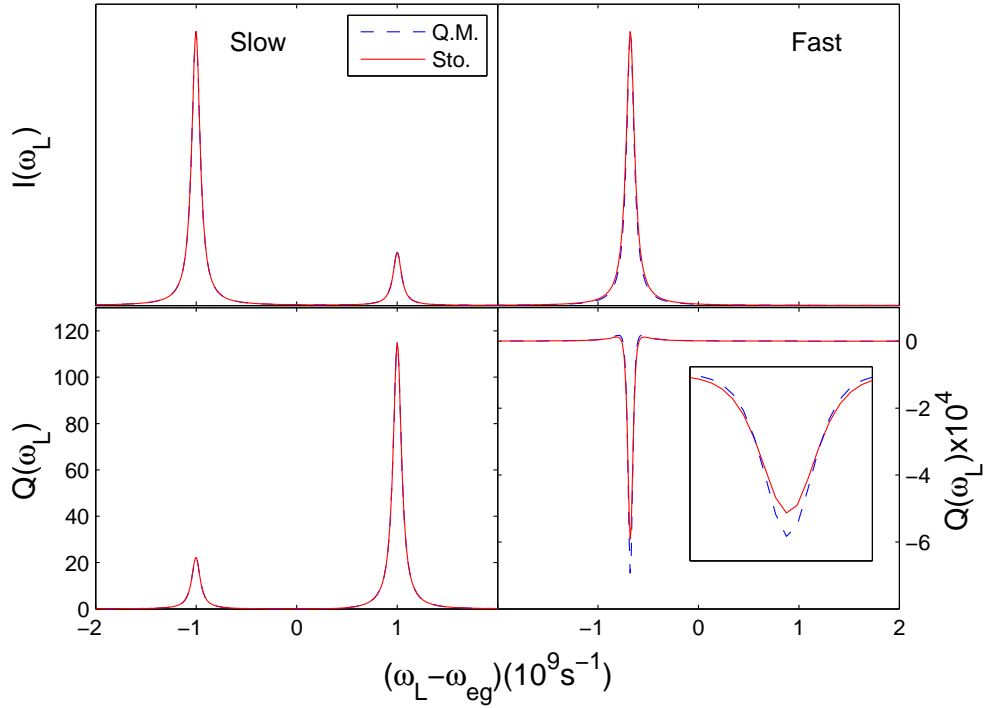, width=5.25in}
\caption{Absorption line shapes and Mandel's  $Q$ parameter spectrum
in the limit of weak coupling between the chromophore and TLS.  Lineshapes
are presented in arbitrary units.
Left and right halves correspond to slow and fast modulations respectively. 
Physical parameters used in this calculation include 
$\Gamma_0=100Ms^{-1}$, $\Omega_0=1Ms^{-1}$, $T=1.7K$
and quantum model parameters taken from ref.\protect{\cite{brownsilbey}}, namely,
$A=2.8K$, $\alpha=3.75\times 10^{11} nm^{3}s^{-1}$, $r = 5.72 nm$, $J=3\times10^{-4}K$.
For the slow modulation we used $C=3.9\times10^{8}s^{-1}K^{-3}$ while for the fast modulation $C=3.9\times10^{18}s^{-1}K^{-3}$.
Within the stochastic approximation these numbers
translate to (eq. \protect{\ref{eq:nuexp0}})  $\nu=1.02\times 10^9 s^{-1}$ and $E=2.8K$. The upward
flip rate $R_{\uparrow}=23.5s^{-1}$ for the slow modulation and $2.35\times10^{11}s^{-1}$ for the fast modulation. In the slow modulation, no
discrepancy between quantum and stochastic treatments is found. For the fast modulation the line shape is the same for both quantum and
stochastic treatments while in the Q parameter there is a small difference between the models. The inset focuses on this difference. }
\label{fig:chtls1}
\end{figure}

 ``Weak'' coupling between the chromophore and TLS is dictated by the 
 condition $A \gg \frac{\alpha \eta}{2 r^3} =P$.  Physically, this can result from either
a small coupling constant $\alpha$ or a large distance between the chromophore
and TLS. 
As discussed above, in this case,  results of the quantum model and 
stochastic model should be quite similar (at least for the line shapes \cite{silb}). 
In the left panes of fig. \ref{fig:chtls1}, we present the long time lineshape and 
corresponding $Q$ parameter spectrum for the case of slow TLS modulation 
and weak TLS-chromophore coupling.  The physical constants chosen are detailed in
the figure caption and represent realistic numbers
for an organic dye molecule embedded in an amorphous
host \cite{brownsilbey}.  We compare 
the quantum model with the associated stochastic approximation. 
As expected, the line shapes for the 
two approaches are identical at the resolution of the figure.  The two peaks
represent the two optical transitions with appreciable overlap 
($a\rightarrow c$ and $b \rightarrow d$).  The other transitions are so weak
as to be invisible at this scale. The difference in peak heights
is due to the difference in thermal occupation probabilities for the two
TLS states (which are basically unmodified by chromophore state due
to the small value of $P$ in the quantum model).  Peak shape is Lorentzian with both linewidths 
given by the spontaneous emission rate (full width at half maximum is  $\Gamma_{0}$).  The
TLS flipping is so slow in this case that it contributes negligibly to the linewidths.

The right panes of fig. \ref{fig:chtls1} display similar information to the left, but with parameters
chosen to insure that the TLS flip rate is faster than the difference in transition frequencies, $\nu$. For
simplicity we increased the flip rate by increasing the value of $C$.  While this is physically questionable,
it does provide the only direct means to increase the TLS flip rate while leaving all other behavior
identical.   In this case, the
lineshape consists of only a single peak due to motional narrowing of the optical transition \cite{kubo,silb}. 
As in the slow modulation limit, we find quantitative correspondence between stochastic and quantum
models for the lineshape calculation.  The stochastic model does deviate slightly from the quantum result
in the calculation of the $Q$ parameter.  Though the deviation is slight, it is interesting to note that there
are cases where the stochastic model is perfect for lineshapes, yet imperfect for higher order statistics.
All in all though,  for weak coupling, the stochastic approximation is seen to perform well both at slow and fast
TLS modulation rates.  We note that in the limiting cases of slow and fast modulation displayed here,
the observed spectra can also be predicted on the basis of the physical approximations introduced
in ref. \cite{zheng4}.

\subsubsection{Strong coupling between chromophore and TLS.}

\begin{figure}
\epsfig{file=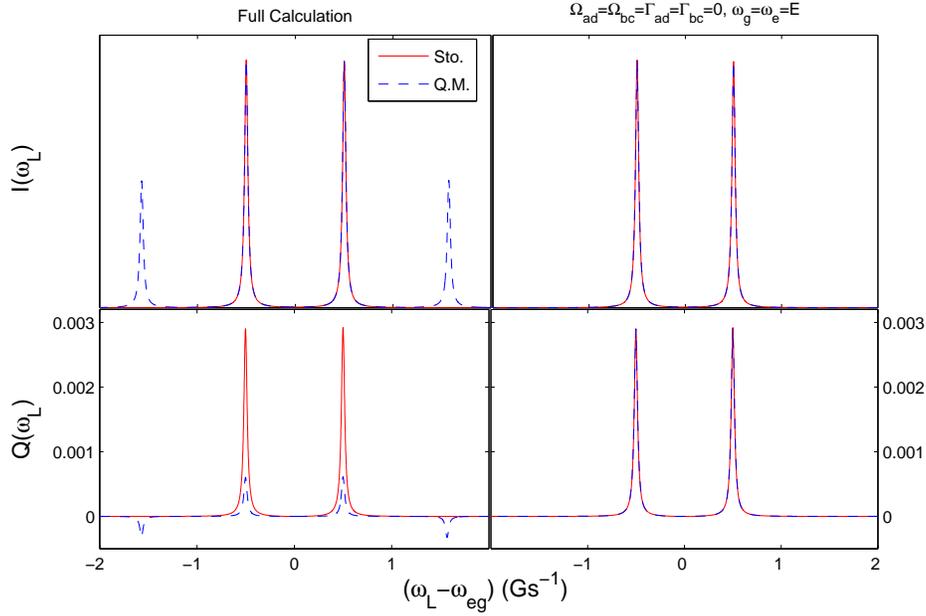, width=5.25in}
\caption{Left panes: The line shape and Mandel's $Q$ parameter spectrum for slow TLS modulation 
with strong coupling between the chromophore and TLS.  
Due to the strong coupling, $d \rightarrow a$ and $c \rightarrow b$ transitions are significant within a 
fully quantum framework and result in two additional peaks relative to weak coupling results.  The
stochastic approach completely misses these additional spectral lines and fares poorly in reproducing
the magnitude of peaks in the $Q$ spectrum. 
The plots correspond to quantum model parameters: 
$\Gamma_{0}=40Ms^{-1}$, $\Omega_{0}=0.1Ms^{-1}$, $T=1.7K$, $A=0.006K$, $J=0.008K$, $C=3.9\times10^{8}K^{-3}s^{-1}$,
$\alpha=3.75\times10^{11}nm^3s^{-1}$, $r=5.72nm$.  Corresponding
stochastic parameters are: $\nu = 501 Ms^{-1}$, $ E =0.01K $ and
$ R_{\uparrow}=42307s^{-1} $.  The right panes display that it is possible to reduce the fully quantum
mechanical treatment to the stochastic results by turning off half of the allowed transitions and calculating
Redfield elements in a manner consistent with the stochastic approach (see text).  In other words, it 
is relatively simple to trace the failures of stochastic modeling.}
\label{fig:chtls3}
\end{figure}

``Strong'' coupling is insured by the condition $A \sim P=\frac{\alpha \eta}{2r^3}$. In this case, the quantum model
differs from its associated stochastic approximation in both line shape and 
Mandel's $Q$ parameter.
The left panes of fig. \ref{fig:chtls3} display results for the strong coupling and slow modulation parameter regime of 
both the quantum and stochastic dynamic treatments. In contrast to
our earlier example, strong coupling now implies that transitions
between states $d \rightarrow a$ and $c \rightarrow b$ are important and occur with some finite probability within
the fully quantum treatment. Since peak widths are smaller than interpeak spacing,  peaks corresponding to all four 
possible transitions are clearly visible in the quantum mechanical modeling.
The relative height of the two central peaks in the line shape are (as in the previous example) related to TLS
thermal occupation probabilities.  Since $E\ll kT$ for the chosen parameters, both central peaks have effectively the 
same height.  The intensity of the outer two peaks is predicted based on the probability to excite an ``off diagonal''
transition ($a\rightarrow d$ or $b \rightarrow c$) relative to diagonal transitions.  Mathematically
this probability is dictated by the square of the Rabi frequency for the transition in question.  Equivalently (see eqs.
\ref{eq:gam_mn} and \ref{eq:omg_mn}),  the ratio of the left two peaks or the right two peaks is
predicted to be $\Gamma_{db}/\Gamma_{da}$ (1.94 for the case shown), which agrees with the numerical
results. It is obvious that the stochastic approximation predicts a very different line shape and $Q$
parameter since it doesn't account for the transitions $d \rightarrow a$ and $c \rightarrow b$.  While one could
argue that the stochastic model does do a good job in predicting that portion of the absorption lineshape
which it is capable of reproducing (the center two peaks), even the center two peaks are clearly off in magnitude
for the $Q$ parameter.  The stochastic model fares very poorly in this parameter regime (strong coupling, slow modulation).

The failure of the stochastic model in this case was predictable and we can trace its origins back to failures
to reproduce the full system dynamics in a realistic manner.  The right panes of 
fig. \ref{fig:chtls3} are meant to display that we understand exactly where these failures are coming from.
These panes actually display two different cases (although they overlap so only a single line is visible):
the stochastic calculation from the left panes and a modified quantum calculation where the evolution
operator was altered such that all non-diagonal transitions were turned off 
($\Omega_{ad}=\Omega_{bc}=\Gamma_{ad}=\Gamma_{bc}=0$ and $\Omega_{ac}=\Omega_{bd}=\Omega_0$
and $\Gamma_{ac}=\Gamma_{bd}=\Gamma_{0}$) and all Redfield elements were calculated
assuming that $\omega_{g}=\omega_{e}=E$.  While these two changes do not fully reduce the quantum
calculation to the stochastic treatment from a mathematical standpoint,  the physical basis is clear.  The
alterations explicitly remove the non-diagonal transitions that the stochastic model necessarily misses and it evaluates
the TLS jump rates in the same approximation inherent to the stochastic approach.  There are more subtle
effects within the Redfield treatment (as in the evolution of coherences) 
so that our ad hoc alterations do not fully limit to a stochastic model, however these effects clearly 
do not contribute to the lineshape and $Q$ spectrum calculations.  The primary problem with a stochastic
model in predicting photon counting observables is in the loss of   ``off-diagonal'' nuclear transitions and
incorrect estimation of relaxation rates.

\begin{figure}
\epsfig{file=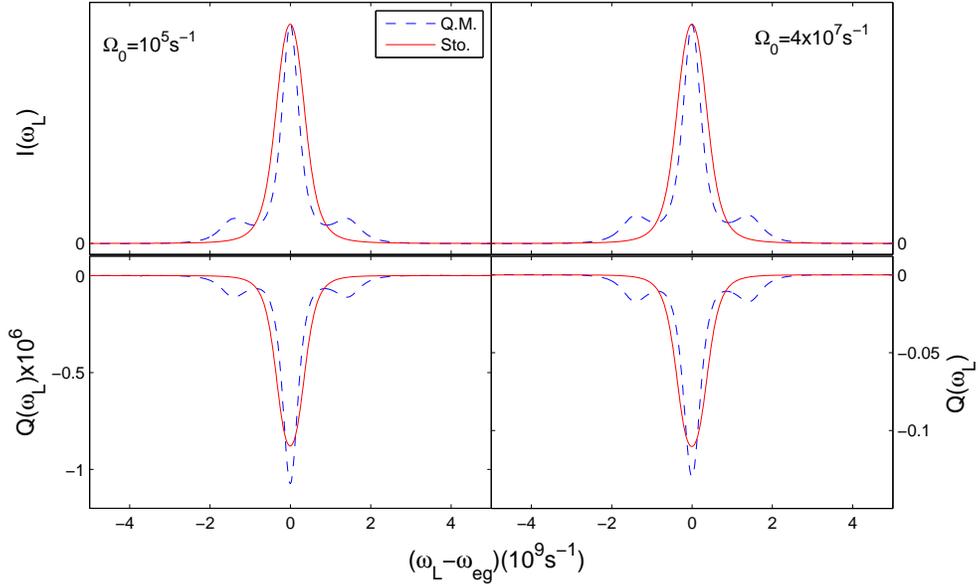, width=5.25in}
\caption{Line shape and Mandel's $Q$ parameter for intermediate TLS modulation rate, 
with ``strong'' coupling between the chromophore and TLS.
The quantum model parameters are the same as in fig. \ref{fig:chtls3} except for the coupling constant which is modified to $C=3.9\times10^{12}K^{-3}s^{-1}$, corresponding to upward
flip rate $ R_{\uparrow}=4.23\times10^{8}s^{-1} $ in the stochastic model.
In the left panes the Rabi frequency coefficient is $\Omega_{0}=10^{5}s^{-1}$, while in the right panes $\Omega_{0}=\Gamma_0=40Ms^{-1}$.
Comparison of the left and right panes shows that antibunching increases as excitation and emission rates 
become comparable.}
\label{fig:chtls4}
\end{figure}

In fig. \ref{fig:chtls4} we show two cases of reasonably fast modulation speed and
strong coupling; the difference between left and right panes is quantitative (see the figure
axes for $Q$) and is intended to display the fact that you can tune the $Q$ parameter
by adjusting field strengths.  For a simple two level chromophore, antibunching is 
maximized when excitation and emission rates are equalized \cite{frank_rev} and
a similar effect is seen here.  Although both quantum and stochastic models will
eventually narrow into a single peak for high enough flip rates, it is interesting to
see in this intermediate regime that the stochastic model has already narrowed,
while the quantum picture retains a more complex structure.  This structure is visible
in both the lineshape and $Q$ parameter calculations.

\subsubsection{Emission spectra}

\begin{figure}
\epsfig{file=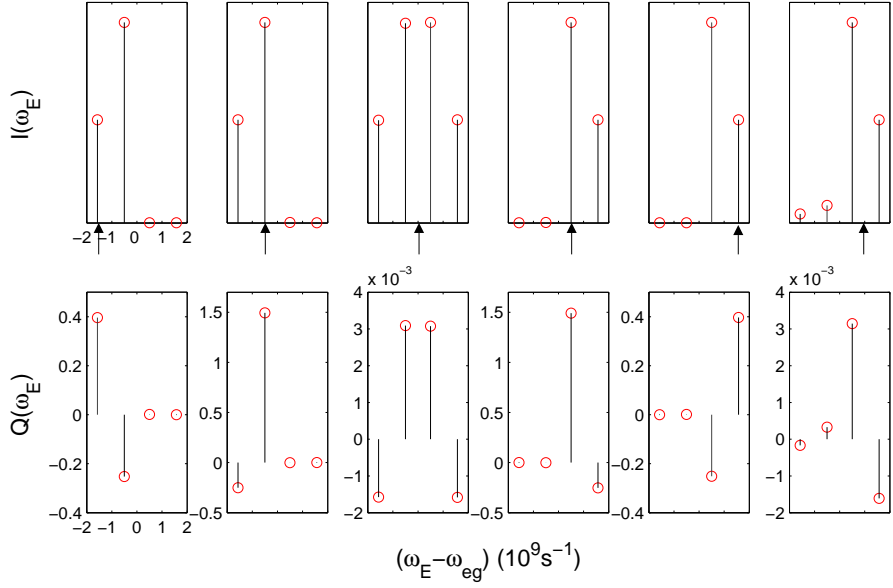, width=5.25in}
\caption{Emission lineshapes and the Mandel's parameter $Q$ for 
slow modulation limit with ``strong'' coupling between
the chromophore and TLS. The excitation laser frequencies are marked in the figure using
``$\uparrow$''. The excitation frequencies, from left to right, are
$\omega_{eg}+\omega_{cb}$, $\omega_{eg}+\omega_{ca}$, $\omega_{eg}$, $\omega_{eg}+\omega_{db}$, 
$\omega_{eg}+\omega_{da}$, and $\omega_{eg}+0.6 \omega_{da}$
(see fig. \ref{fig:modeltls}).
The spontaneous emission rate and the Rabi frequency are 
$\Gamma_{0}=40Ms^{-1}$, $\Omega_0=4Ms^{-1}$, respectively. 
The quantum model parameters are: $T=1.7K$,
$A=0.006K$, $J=0.008$, $\alpha=3.75\times 10^{11} nm^{-3}s^{-1}$, 
$C=3.9\times10^{8}K^{-3}s^{-1}$ and $r = 5.72nm$.
}
\label{fig:tlsemi}
\end{figure}

 In fig. \ref{fig:tlsemi} we display 
emission line shapes and Mandel's $Q$ parameter spectra for the
same physical parameters selected in fig. \ref{fig:chtls3} (excepting the
Rabi frequency, which was set to provide relatively large magnitudes
of the $Q$ parameter in the anti-bunching regimes).  As discussed previously,
our simulation methodology does not allow for true calculation of emission spectra.
The frequency dependence we obtain is resolved solely on the basis of individual 
state to state transitions - we assign all photons emitted for a given transition the 
resonance frequency of that transition.  Hence, the ``lineshapes'' in fig. \ref{fig:tlsemi}
are not broadened by the radiative lifetime of the chromophore or by any other source
and line shifts are not captured.  Physically, the spectra we obtain would match an 
experimental measurement with an instrument unable to resolve frequency differences 
less than the radiative line width.

The multiple panels in both rows of fig. \ref{fig:tlsemi} reflect different laser exciting frequencies.
Four different resonant excitations corresponding to all possible transitions and two off resonant
frequencies are considered.  Clearly, there is a strong dependence in the emission spectra on
the exciting frequency.  This is expected since TLS dynamics are slow enough in this problem that the
TLS does typically not have a chance to relax to equilibrium while the chromophore is excited.  Resonant
excitation to state $c$, regardless of which ground state ($a$ or $b$) the transition starts from
results in the same emission line shape (left two panes of the top row of fig. \ref{fig:tlsemi}).  The
relative peak heights simply reflect Condon overlaps in the spontaneous emission process 
from state $c$ back to $a$ or $b$.  These overlaps don't care how state $c$ was excited and
generate identical emission spectra regardless of which resonant transition is excited.  Similar
arguments explain the right three panes of the top row of fig. \ref{fig:tlsemi}.  All three 
excitation frequencies result in the occupation of state $d$ and the emission lineshapes are
insensitive to details of the excitation beyond this fact - even when the excitation is off resonance
with either $a\rightarrow d$ or $b \rightarrow d$ transitions.  When an off resonant excitation is
considered that has equal probability to excite to either $c$ or $d$, the emission lineshapes reflect
a symmetric combination of the previously discussed cases (third pane of the top row of the 
figure).

In contrast to the lineshapes, $Q$ parameter spectra are highly sensitive to excitation frequency
(bottom row of fig. \ref{fig:tlsemi}).  The basis for this effect is quite simple.  When photons are counted
at the same frequency of the exciting laser we expect to see photon bunching.  For example, looking
at the leftmost peak in the leftmost pane of the bottom row we excite $b\rightarrow c$ transitions
and monitor $c \rightarrow b$ emissions.  Photons are repeatedly ejected as this cycle repeats until spontaneous
emission induces a $c \rightarrow a$ transition (or the TLS flips), at which point the system is off resonance and
has to wait for a TLS flip to return the system to the excitable state $b$.  The interspersion of
bright and dark intervals leads to bunching phenomena and a positive $Q$ parameter.  In
contrast, when excitation does not correspond to the monitored transition (second peak from
left in the leftmost pane) a three state cycle repeatedly occurs ($b \rightarrow c \rightarrow a \rightarrow b \ldots$
or a similar variant) as photons are detected.  There is no jumping between periods of ``bright'' or ``dark'' since the
pathway for repeated photon emission necessarily involves both TLS and radiative/excitation dynamics.  The
chosen timescales in this example insure that no single rate is limiting over all others in this cycling 
process and antibunching results (if a single timescale were completely dominant we would expect $Q=0$).  
Similar arguments can be applied to the remaining panes of the $Q$ parameter spectrum.  This example
makes a clear case for measurement of higher order photon counting moments.  Different aspects of 
system dynamics are captured in the measurement of the $Q$ parameter beyond what is seen in
simple lineshape statistics.  Furthermore, examination of the emission statistics provides a more detailed
measure than possible solely on the basis of absorption statistics.

\section{A chromophore with nuclear vibrations coupled to an harmonic bath}
\label{sec:harmonic}
\subsection{Model description}

\begin{figure}
\epsfig{file=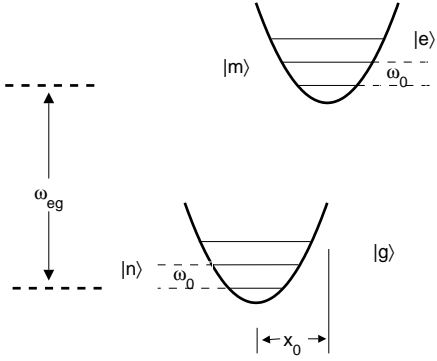, width=3.25in}
\caption{ Schematic description of a system with two electronic levels and single harmonic vibrational mode. }
\label{fig:harmodel}
\end{figure}

As a more complex example of multilevel quantum dynamics we consider the case of a 
chromophore with an harmonic vibrational degree of freedom.  Coupled to this vibrational coordinate
is a bath modeled by an ensemble of harmonic oscillators.  Such models are standard in the 
treatment of molecular spectroscopy \cite{mukbook}, but have seen little prior use in the
treatment of photon statistics.
Within the Born-Oppenheimer approximation, the Hamiltonians of the
chromophore in its 
electronic ground, $|g \rangle$, and excited,  $|e \rangle$ ,
states are taken to be
\begin{eqnarray}
\label{eq:harham}
H_g & = & \frac{1}{2} \hbar \omega_{0} [ P^2+X^2 ],   \\\nonumber
H_e & = & \hbar\omega_{eg}+ \frac{1}{2} \hbar \omega_{0} [ P^2+(X-X_0)^2 ], 
\end{eqnarray}
where $X$ and $P$ are related to the nuclear position coordinate $x$ and momentum $p$ by
\begin{eqnarray}
X=\sqrt{\frac{m \omega_{0}}{\hbar}}x,    \\\nonumber
P=\frac{1}{m \omega_{0}\hbar}p.
\end{eqnarray}
The vibrational coordinate thus has frequency $\omega_{0}$ and
$\hbar \omega_{eg}$ is the excitation energy for the $0-0$ transition.
$x_0$ is the shift in equilibrium position of the nuclear coordinate between
excited and ground states (see fig. \ref{fig:harmodel}).
The interaction with the thermal bath is assumed to be linear in both $X$
and bath coordinates $X_i$, i.e.,
\begin{equation}
\label{eq:harcoup}
V= CX \sum_j X_j 
\end{equation}
where $C$ is a constant specifying the interaction strength between system
and bath. The harmonic bath Hamiltonian is
\begin{equation}
\label{eq:bath}
H_B = \sum_j \frac{1}{2} \hbar \omega_j \left( P_j^2 + X_j^2 \right).
\end{equation}
The above definitions of $H_g$, $H_e$, $V$ and $H_B$ provide all necessary
information to proceed directly with the calculation of $\mathbb{L}$ and related
quantities as detailed in section \ref{sec:theory}.  We make a few brief comments
related to the calculation of Redfield elements below in order to clarify our notation. 
More detailed presentations can be found elsewhere \cite{blum,mukbook,friesner}.

The linear interaction between bath and system in only capable of effecting transitions between adjacent vibrational
states in the same electronic manifold, i.e. $\ket{n} \rightarrow \ket{n+1}$ or $\ket{n} \rightarrow \ket{n-1}$.
This is seen, by introducing the usual creation and annihilation operators ($a=(X+iP)/\sqrt{2},a^{\dagger}=(X-iP)/\sqrt{2}$) to write the interaction matrix elements between excited state levels in the form
\begin{equation}
V_{n_{e},n'_{e}}=\frac{1}{2}C[\sqrt{n'_{e}}\delta_{n_{e},n'_{e}-1}+\sqrt{n_{e}}\delta_{n_{e},n'_{e}+1}]\sum_j(a_j+a^{\dagger}_j)
\end{equation}
and similarly for the ground state.  The creation and annihilation operators only allow for adjacent transitions
as indicated by the above delta functions.  The bath properties
\begin{eqnarray}
a_j(t)&=&e^{-i\omega_j t}a_j(0)   \\\nonumber
a_j^{\dagger}(t)&=&e^{i\omega_j t}a_j^{\dagger}(0)  \\\nonumber
\langle a_ja_j^{\dagger} \rangle_b &=& (1-e^{-\beta \hbar \omega_j})^{-1}  \\\nonumber
\langle a_j^{\dagger}a_j \rangle_b &=& e^{-\beta \hbar \omega_j}(1-e^{-\beta \hbar\omega_j})^{-1}.
\end{eqnarray}
are used to evaluate all correlation functions associated with the Redfield matrix calculation.
In this model the interaction matrix $V$ is explicitly real leading to a slightly simplified
calculation for the Redfield matrix
\begin{equation}
\mathbb{R}_{mnpq}=t_{pmnq}+t_{qnmp}-\delta_{mp}\sum_{r}t_{nrrq}-\delta_{nq}\sum_{r}t_{mrrp},
\end{equation}
where
\begin{equation}
t_{pmnq}=\frac{1}{2\hbar^2}\int_{-\infty}^{\infty} d\tau e^{i\omega_{qn} \tau}\langle \hat{V}_{pm}(\tau)\hat{V}_{nq}(0) \rangle_{b}.
\end{equation}
$t_{pmnq}$ is non-zero only if both of the pairs $(p,m)$ and $(n,q)$ involve states in the same
electronic manifold.
The integration can be carried out and yields
\begin{equation}
t_{pmnq}=R_0[\sqrt{m}\delta_{p,m-1}+\sqrt{p}\delta_{p,m+1}][\sqrt{n}\delta_{q,n-1}+\sqrt{q}\delta_{q,n+1}]
\sum_j \frac{\delta(\omega_j+\omega_{qn})+e^{-\beta \hbar \omega_j} \delta(\omega_j-\omega_{qn})}{1-e^{-\beta \hbar \omega_j}}.
\end{equation}
Unlike the TLS model, in this case every allowable $\omega_{qn}$ is exactly the same and is equal to
$\omega_{0}$.  This is due to the equality of spacing between levels in the harmonic oscillator model and
the form of $V$ which only allows for adjacent transitions.  Thus, the density of bath states is not important in calculating the Redfield matrix elements in this case and only a single constant $R_0$ enters into the Redfield description 
as a measure of coupling between system and bath. 
For example, elements of the form $\mathbb{R}_{nn;n+1n+1}$ are given in our notation by
\begin{equation}
\mathbb{R}_{nn;n+1n+1}=2R_0(n+1)(1-e^{-\beta \hbar \omega_0})^{-1}.
\end{equation}
Since this element reflects the rate of transition from harmonic oscillator state $\ket{n+1}$ to $\ket{n}$, it is clear that $R_0$ is closely related to the relaxation rate of our vibrational coordinate.
\begin{figure}
\epsfig{file=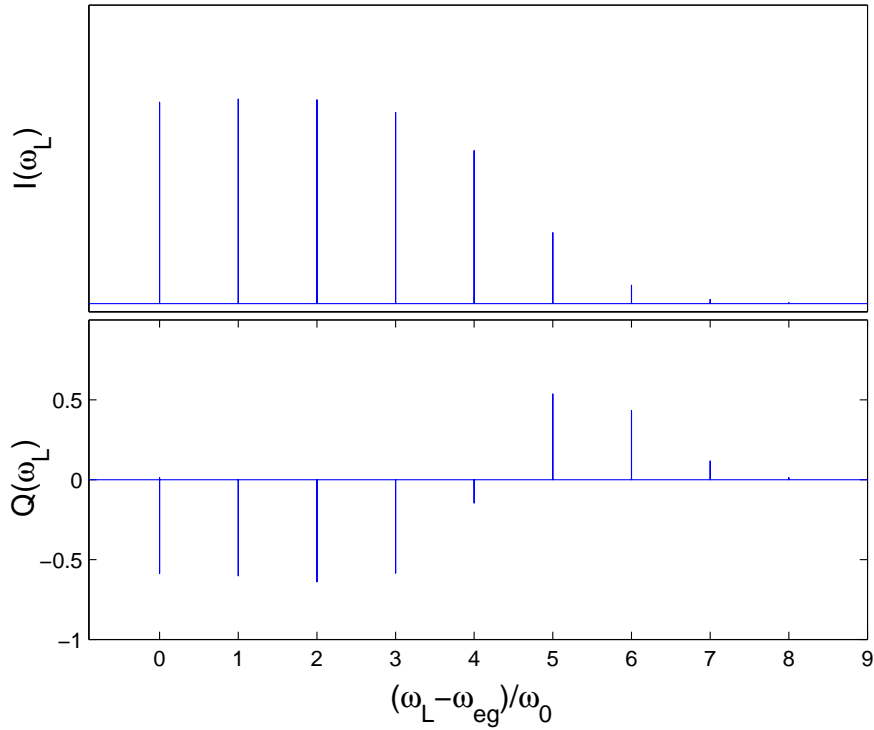, width=5.25in}
\caption{ The line shape and the Mandel's $Q$ parameter
spectrum as a function of exciting laser frequency for a chromophore
with an harmonic vibrational coordinate.
The spontaneous emission rate and Rabi frequency are 
$\Gamma_0=\Omega_0=10^8s^{-1}$ and
the coupling strength is $R_0=10^7$.  Physical parameters
specific to the chromophore are detailed in the text.}
\label{fig:I2nq}
\end{figure}

\subsection{Numerical results}

In the following calculations we choose physical parameters specifying the chromophore to be $\omega_0=3.77 \times 10^{13}s^{-1}$, $x_0=0.11$\AA, $ m=10^5 m_e $ ($ m_e $ is the electron mass) and $ T=10K $ 
(The energy difference between neighboring levels of the harmonic oscillator $\hbar\omega_0$
corresponds to temperature of $287.75K$).  While these numbers are suggestive of a heavy diatomic
molecule (like $I_2$) in a low temperature matrix we have not made a serious attempt to connect these
calculations with physical systems.  Rather, we have chosen $x_0$ to provide Condon overlaps that are
close to vertical, while still insuring finite probability for transitions up to $0-6$.  We have also set the temperature
somewhat arbitrarily while we will freely adjust $R_0$ in the following examples to meet our needs in displaying
various phenomena.  The Redfield approach we employ is necessarily limited to a finite number of states due
to numerical considerations.  We can not solve the equations for $N = \infty$. In the numerics presented here 
we used 10 levels in each of the electronic states ($n=0$ to $n=9$).  It was verified that altering the number
of vibrational states to include more levels did not change any results at the resolution of the presented
figures.  We note that the size of $\mathbb{L}$ for these calculations is $400\times 400$.  Using the methods
of sec. \ref{sec:practical} requires only diagonalization of this matrix, which is a simple task for modern 
computers.
 
\subsubsection{ Weak coupling between system and bath ($R_0$ small) case.}

The case of weak coupling corresponds to slow vibrational relaxation. 
In fig. \ref{fig:I2nq} we show the line shape and the $Q$ parameter for a case in
which the relaxation rate is slower than all other rates in the problem including the spontaneous emission rate, Rabi frequency and oscillator frequency.  This leads to
non-thermal distributions of vibrational levels within both electronic manifolds at steady state since the
system is unable to fully relax between subsequent photon emission/absorption events.  Interestingly, the 
variation of these steady states with excitation frequency and the variation of Condon overlaps between
the various transitions leads to $Q$ parameter values spanning a range of positive and negative values
depending upon the excitation frequency. It should be noted that although the spectra appear to have
only been evaluated at  the various allowed resonance frequencies, this is not the case.  It is simply 
the case that the radiative linewidths are very much narrower than discernible at the resolution of the
figure.  

\begin{figure}
\epsfig{file=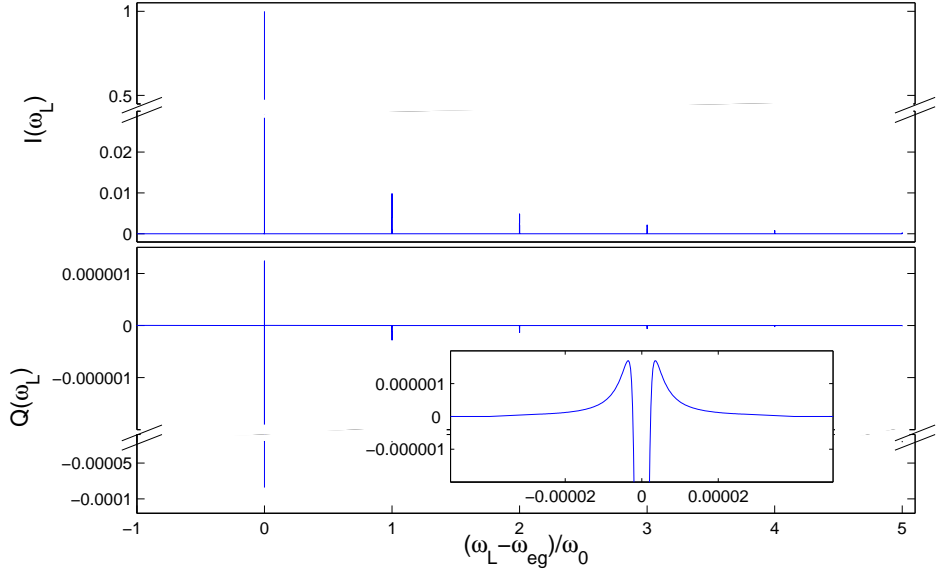, width=5.25in}
\caption{Similar to fig. \protect{\ref{fig:I2nq}}, but with $R_0=10^{10}s^{-1}$, $\Gamma_0=10^8s^{-1}$ and $\Omega_0=10^6$. This system is in the linear response regime.  The inset
shows the variation of $Q$ in the vicinity of $\omega_L=\omega_{eg}$. }
\label{fig:Abs}
\end{figure}

Fig. \ref{fig:Abs} shows the line shape and $Q$ parameter for a case in which the relaxation is slow relative to the 
harmonic oscillation frequency $\omega_0$, but  is
faster than the spontaneous emission rate and the Rabi frequency. In this case the 
relative amount of power absorbed by each possible transition is expected to agree with linear response predictions since the vibrational state of the chromophore should almost always be in the relaxed ($n=0$)
state without significant perturbation by the relatively weak coupling to the field.  Linear response theory
predicts that the strength of each transition is due to the Condon overlap between $n=0$ in the 
ground state (remember $kT\ll \hbar \omega_0$ in this model) and the various excited states.  The displayed 
lineshapes appear to contradict this prediction, most clearly due to the very tall zero phonon peak at $\omega_L = \omega_{eg}$
relative to the other peaks.  However, the height of this line is due to the fact that this transition is not broadened
by non-radiative processes as are the remaining transitions.  The linewidth of the $0-0$ line is approximately
equal to $\Gamma_0$ whereas the other widths are dominated by non-radiative decay on the order of
$R_0$ and are 100 times wider.  The relevant quantities to compare with linear response results are the 
intensities of each transition integrated over the local vicinity of the transition.  In fig. \ref{fig:emi1}
we display such integrated absorption peaks alongside emission lines (discussed below).  These
integrated lines show perfect agreement with linear response results with relative intensities directly
proportional to the square of nuclear overlap.

One interesting point to note about the $Q$ parameter in these calculations is that it undergoes rapid
variation with excitation frequency in the vicinity of the $0-0$ line.  While this behavior does not seem
amenable to simple explanation, it has been observed previously in simpler models both
numerically \cite{zheng3} and analytically \cite{bark1}.  It should also be emphasized that the magnitude of 
$Q$ is largely due to the ratio between $\Gamma_0$ and $\Omega_0$ as seen in fig. \ref{fig:chtls4}.
Here this ratio is large, leading to small negative $Q$ values.  Smaller ratios lead to larger magnitudes
of $Q$ (when Q is negative).

\subsubsection{Strong coupling between the system bath ($R_0$ not small) case:}
\begin{figure}
\epsfig{file=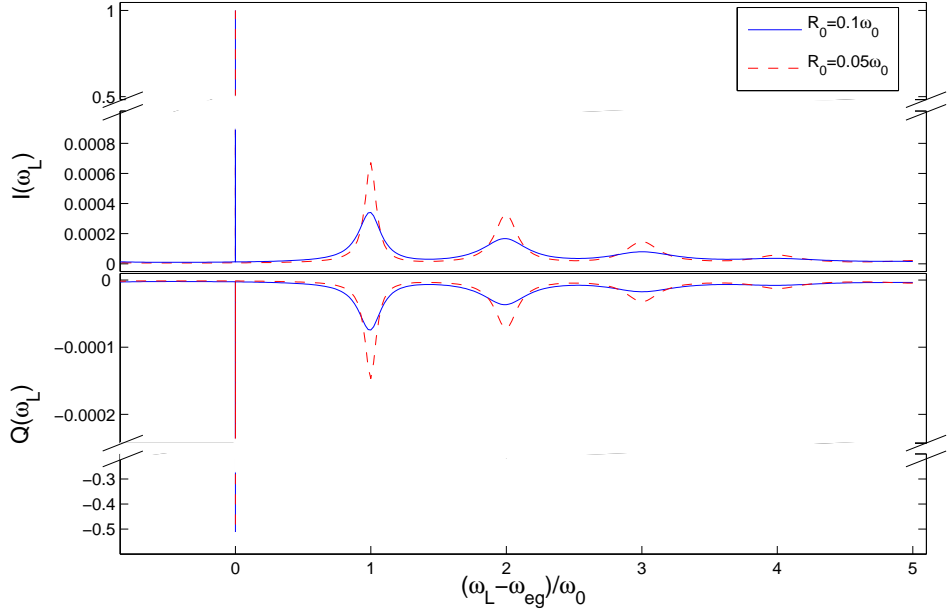, width=5.25in}
\caption{Similar to fig. \protect{\ref{fig:I2nq}}, but with  $\Gamma_0=\Omega_0=10^9s^{-1}$ and 
$R_0=0.1\omega_0$ (solid line)  and $R_0=0.05\omega_0$ (dashed line). Note that in this case the width of the peak at $0$ is much smaller than
the width of the other peaks, since it does not depend on $R_0$.  Peak widths are given by the non-radiative
lifetime of the various states for all other transitions.}
\label{fig:I2n}
\end{figure}

 An example of fast relaxation, with $R_0$ on the order of $\omega_0$, is shown in fig. \ref{fig:I2n}.
In this case the width of the peaks is of the same order as the distance between the peaks and line shape 
is clearly not a series of thin sticks as in previous examples. Note that since the peak at $\omega_L=\omega_{eg}$ does not involve any thermal relaxation it is independent of $R_0$.  The width 
of this peak is still specified by the spontaneous emission rate, which is orders of magnitude lower than
the remaining peak widths (on the order of $R_0$) leading to its very large height.  In this plot we have 
chosen identical values for $\Gamma_0$ and $\Omega_0$, which leads to sizable negative $Q$ values
for the $0-0$ line.

\subsubsection{Emission spectroscopy }
\begin{figure}
\epsfig{file=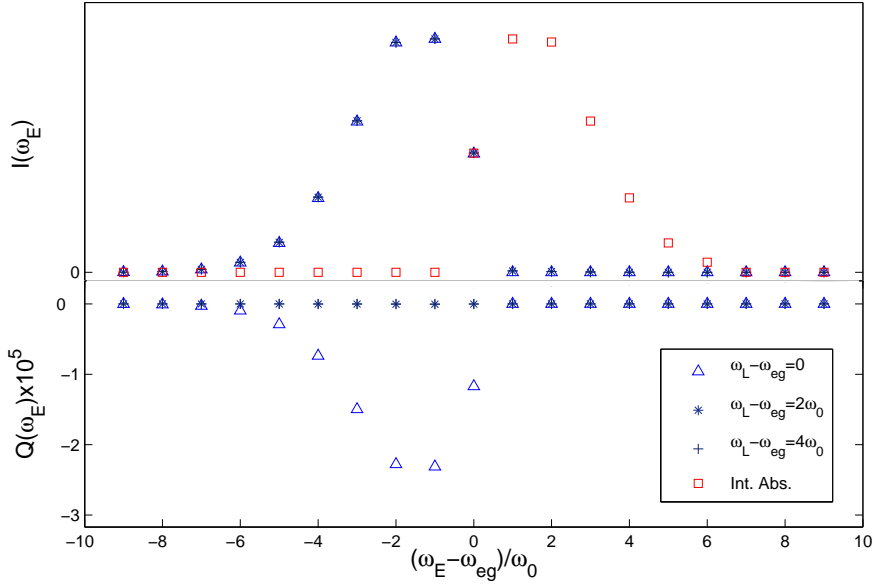, width=5.25in}
\caption{The emission line shape and emission $Q$ spectrum for parameters reflecting the
linear response regime.  Three different excitation frequencies are considered as noted in the legend.
The chosen physical parameters parallel those of \protect{\ref{fig:Abs}}. Since the system behaves
in accord with linear response,
the emission line shapes are the same for all excitation frequencies and also in agreement (mirror image) 
with the integrated absorption spectrum.}
\label{fig:emi1}
\end{figure}

In fig. \ref{fig:emi1} we show the emission line shape and $Q$ parameter spectra for parameters appropriate
to the linear response regime (identical parameters to fig. \ref{fig:Abs}). It is shown that in this case the line shape is the same for all excitation frequencies (in the
fig we show $\omega_L-\omega_{eg}=0,2\omega_0,4\omega_0$). It is also shown that integration of the absorption spectrum over the individual transition linewidths provides a mirror image of the emission line shape as expected
in the linear response regime.  Recall that our emission line shapes are sensitive only to individual transitions,
so the emission spectra are automatically of the ``integrated'' type and comparison between emission and integrated
absorption is completely natural.  While emission lineshapes are insensitive to excitation frequency in
this regime, the $Q$ parameter exhibits strong dependence on excitation frequency.

\begin{figure}
\epsfig{file=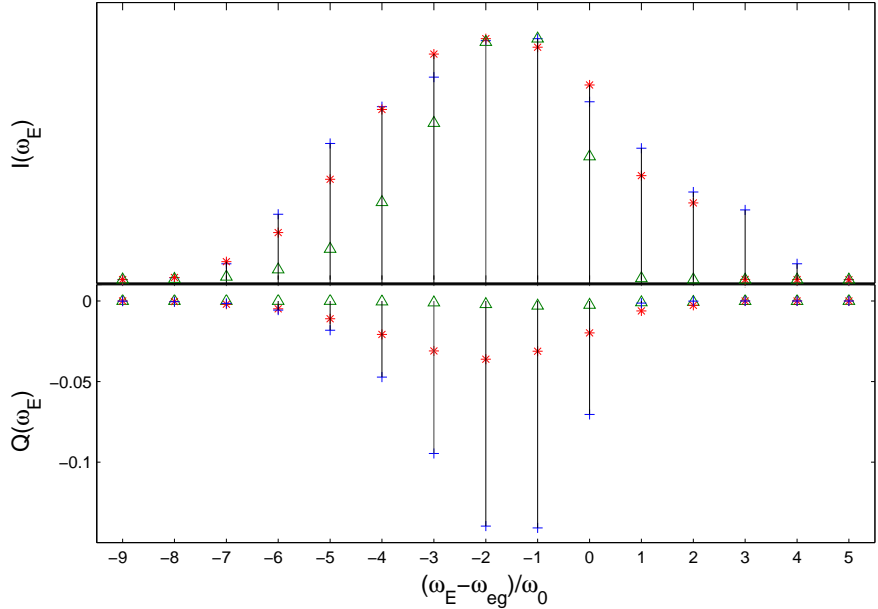, width=5.25in}
\caption{Emission line shape and $Q$ spectrum for parameters outside the linear response limit. 
$R_0=\Gamma_0=\Omega_0=10^{8}s^{-1}$. 
The exciting fields are as in fig. \ref{fig:emi1}. 
In this case the linear response approximation is not valid anymore and both the line shape and $Q$ spectra differ with excitation frequency.}
\label{fig:emi2}
\end{figure}

Fig. \ref{fig:emi2} shows the emission line shape and $Q$ parameter for stronger driving fields  and slower relaxation rates than present in fig. \ref{fig:emi1}.  The system is no longer in the linear response regime and 
line shapes differ for different excitation frequencies.
The parameters were chosen to equalize all relevant physical timescales, demonstrating that there is no
simple relationship possible between excitation frequency, emission lineshape and emission $Q$ parameter
possible in general.

\section{conclusion}
\label{sec:conc}
We have introduced a practical framework for the calculation of photon counting statistics
in quantum systems with multiple levels and dissipative coupling to a thermal environment.  The
present scheme generalizes previous work by extending the treatment of chromophore
dynamics beyond the stochastic models historically applied to single molecule spectroscopy.
Our model calculations for TLS dynamics explicitly demonstrate some of the failings of traditional 
stochastic modeling.  In the case of harmonic vibrations, use of a stochastic model is even more
suspect since all quantization of the vibrational coordinate will be lost.  Although one could 
envision more elaborate kinetic schemes in an attempt to model these systems, it 
seems more straightforward to simply treat the dynamics
correctly, quantum mechanically, from the outset.  The methods presented here provide a prescription
to do this.

We acknowledge that there is an unfortunate amount of machinery behind the calculations that
we have presented here, however it is important to stress that 90\% of this machinery is associated
with the implementation of the Redfield formalism (calculation of the matrix $\mathbb{L}$ in our
notation).   Eqs. \ref{eq:newmethod} are very simply applied
once $\mathbb{L}$ is given; simply diagonalize the matrix and perform a few simple matrix multiplications
as implied by the formulae.  The generating function approach, while necessarily 
encumbered by the usual difficulties in simulating dissipative quantum systems, adds no new significant
conceptual or numerical problems.  Photon counting statistics are therefore readily available at no
more expense than normally expected for calculation of density matrix dynamics.  This remarkable fact
seems to be the strongest point in support of the generating function methodology.

Several of our calculations have presented results for emission spectra and the
corresponding $Q$ parameter quantities.  Although such measurements are not yet
within the capabilities of experiment, we believe that a strong case can be made for
the development of single molecule detectors with spectral resolution.  It is clear from
our model calculations that emission spectroscopy provides a different and (when combined
with absorption spectroscopy) more revealing signature of chromophore dynamics than obtainable
from absorption alone.  This is not surprising, but the present study is (to our knowledge) the 
first to demonstrate this fact explicitly.  As we have repeatedly stated, the present scheme for
emission spectroscopy is sensitive only to molecular transitions and not directly to emission 
frequency.  Emission frequency is assumed to be on resonance with specific transitions.  While
this approach works well in the limit of weak coupling to the environment, stronger coupling
invariably leads to level shifts, motional narrowing as associated complications.  A general
and practical formulation of true emission photon counting statistics has yet to be developed.

\section*{Acknowledgment}
This research was supported in part by the 
Research Corporation and the National Science
Foundation (CHE-0349196, CHE-0321368).  F.B. is an Alfred P. Sloan
Research Fellow.


\begin{thebibliography}{10}

\bibitem{moerner}
Moerner,~W.~E.;\ \ Kador,~L. Optical Detection and Spectroscopy of Single
  Molecules In a Solid. \textit{Phys. Rev. Lett.} \textbf{1989,} \textsl{62,}
  2535--2538.

\bibitem{orrit}
Orrit,~M.;\ \ Bernard,~J. Single Pentacene Molecules Detected by Fluorescence
  Excitation in a p-terphenyl Crystal. \textit{Phys. Rev. Lett.} \textbf{1990,}
  \textsl{65,} 2716--2719.

\bibitem{moernerrev}
Moerner,~W.~E.;\ \ Orrit,~M. Illuminating sigle molecules in condensed matter.
  \textit{Science} \textbf{1999,} \textsl{283,} 1670--1676.

\bibitem{wildrev}
Plakhotnik,~T.;\ \ Donley,~E.~A.;\ \ Wild,~U.~P. Single-molecule spectroscopy.
  \textit{Ann. Rev. Phys. Chem.} \textbf{1997,} \textsl{49,} 181--212.

\bibitem{xie}
Lu,~H.~P.;\ \ Xun,~L.;\ \ Xie,~X.~S. Single-Molecule Enzymatic Dynamics.
  \textit{Science} \textbf{1998,} \textsl{282,} 1877--1882.

\bibitem{xie_rev}
Xie,~X.~S. Single-molecule approach to dispersed kinetics and dynamic disorder:
  Probing conformational fluctuation and enzymatic dynamics. \textit{J. Chem.
  Phys.} \textbf{2002,} \textsl{117,} 11024--11032.

\bibitem{chu}
Zhuang,~X.;\ \ Bartley,~L.~E.;\ \ Babcock,~H.~P.;\ \ Russell,~R.;\ \ Ha,~T.;\ \
  Herschlag,~D.;\ \ Chu,~S. A Single-Molecule Study of RNA Catalysis and
  Folding. \textit{Science} \textbf{2000,} \textsl{288,} 2048--2051.

\bibitem{sweiss}
Weiss,~S. Fluorescence Spectroscopy of Single Biomolecules. \textit{Science}
  \textbf{1999,} \textsl{283,} 1676--1683.

\bibitem{silbrev}
Jung,~Y.;\ \ Barkai,~E.;\ \ Silbey,~R.~J. Current Status in Single Molecule
  Spectroscopy: Theoretical Aspects. \textit{J. Chem. Phys.} \textbf{2002,}
  \textsl{117,} 10980--10995.

\bibitem{speculate}
Orrit,~M. Single Molecule Spectroscopy: The road ahead. \textit{J. Chem. Phys.}
  \textbf{2002,} \textsl{117,} 10938--10946.

\bibitem{cpc_rev}
Bohmer,~M.;\ \ Enderlein,~J. Fluorescence Spectroscopy of Single Molecules
  under Ambient Conditions: Methodology and Technology. \textit{ChemPhysChem}
  \textbf{2003,} \textsl{4,} 793--808.

\bibitem{orrit_rev}
Lippitz,~M.;\ \ Kulzer,~F.;\ \ Orrit,~M. Statistical Evaluation of Single
  Nano-Object Fluorescence. \textit{ChemPhysChem} \textbf{2005,} \textsl{6,}
  770--789.

\bibitem{silbey_rev}
Jung,~Y.;\ \ Barkai,~E.;\ \ Silbey,~R.~J. Current status in single molecule
  spectroscopy: Theoretical aspects. \textit{J. Chem. Phys.} \textbf{2002,}
  \textsl{117,} 10980--10995.

\bibitem{skinner}
Geva,~E.;\ \ Skinner,~J.~L. Theory of single-molecule optical line-shape
  distributions in low temperature glasses. \textit{J. Phys. Chem. B}
  \textbf{1997,} \textsl{44,} 8920--8932.

\bibitem{wang}
Wang,~J.;\ \ Wolynes,~P. Intermittency of Single Molecule Reaction Dynamics in
  Fluctuating Environments. \textit{Physical Review Letters} \textbf{1995,}
  \textsl{74,} 4317--4320.

\bibitem{wangII}
Wang,~J.;\ \ Wolynes,~P. Intermittency of activated events in single molecules:
  The reaction diffusion description. \textit{J. Chem. Phys.} \textbf{1999,}
  \textsl{110,} 4812--4819.

\bibitem{schenter}
Schenter,~G.~K.;\ \ Lu,~H.~P.;\ \ Xie,~X.~S. Statistical analyses and
  theoretical models of single molecule enzymatic dynamics. \textit{J. Phys.
  Chem. A} \textbf{1999,} \textsl{103,} 10477--10488.

\bibitem{portman}
Portman,~J.~J.;\ \ Wolynes,~P.~G. Complementary variational approximations for
  intermittency and reaction dynamics in fluctuating environments. \textit{J.
  Phys. Chem. A} \textbf{1999,} \textsl{103,} 10602--10610.

\bibitem{agmon}
Agmon,~N. Conformational cycle of a single working enzyme. \textit{J. Phys.
  Chem.} \textbf{2000,} \textsl{104,} 7830--7834.

\bibitem{cao}
Cao,~J. Event-averaged measurements of single-molecule kinetics. \textit{Chem.
  Phys. Lett.} \textbf{2000,} \textsl{327,} 38--44.

\bibitem{caoII}
Yang,~S.;\ \ Cao,~J. Two-event echos in single-molecule kinetics. \textit{J.
  Phys. Chem. B} \textbf{2001,} \textsl{105,} 6536--6549.

\bibitem{mukamel}
Barsegov,~V.;\ \ Chernyak,~V.;\ \ Mukamel,~S. Multitime correlation functions
  for single molecule kinetics. \textit{J. Chem. Phys.} \textbf{2002,}
  \textsl{116,} 4240--4251.

\bibitem{muk3}
Barsegov,~V.;\ \ Mukamel,~S. Multipoint fluorescence quenching-time statistics
  for single molecules with anomalous diffusion. \textit{J. of Phys. Chem. A}
  \textbf{2004,} \textsl{108,} 15--24.

\bibitem{barkai}
Barkai,~E.;\ \ Jung,~Y.;\ \ Silbey,~R. Time-dependent fluctuations in single
  molecule spectroscopy: a generalized Wiener-Khintchine approach.
  \textit{Phys. Rev. Lett.} \textbf{2001,} \textsl{87,} art. no. 207403.

\bibitem{younjoon}
Jung,~Y.;\ \ Barkai,~E.;\ \ Silbey,~R. Lineshape theory and photon counting
  statistics for blinking quantum dots: a Levy walk process. \textit{Chem.
  Phys.} \textbf{2002,} \textsl{284,} 181--194.

\bibitem{brown}
Brown,~F. L.~H. Single molecule statistics with time dependent rates: a
  generating function approach. \textit{Phys. Rev. Lett.} \textbf{2003,}
  \textsl{90,} Art. No. 028302.

\bibitem{wild}
Fleury,~L.;\ \ Segura,~J.-M.;\ \ G.~Zumofen,~B.~H.;\ \ Wild,~U.~P. Nonclassical
  Photon Statistics in Single-Molecule Fluorescence at Room Temperature.
  \textit{Phys. Rev. Lett.} \textbf{2000,} \textsl{84,} 1148--1151.

\bibitem{szabo1}
Gopich,~I.~V.;\ \ Szabo,~A. Statistics of transitions in single molecule
  kinetics. \textit{J. Chem. Phys.} \textbf{2003,} \textsl{118,} 454--455.

\bibitem{szabo2}
Gopich,~I.~V.;\ \ Szabo,~A. Theory of photon statistics in single-molecule
  Forster resonance energy transfer. \textit{J. Chem. Phys.} \textbf{2005,}
  \textsl{122,} Art. No. 014707.

\bibitem{orphir}
Flomenbom,~O.;\ \ Klafter,~J.;\ \ Szabo,~A. What Can One Learn From Two-State
  Single-Molecule Trajectories? \textit{Biophys. J.} \textbf{2005,}
  \textsl{88,} 3780--3783.

\bibitem{plenio}
Plenio,~M.~B.;\ \ Knight,~P.~L. The quantum-jump approach to dissipative
  dynamics in quantum optics. \textit{Rev. Mod. Phys.} \textbf{1998,}
  \textsl{70,} 101--144.

\bibitem{metiu}
Makarov,~D.~E.;\ \ Metiu,~H. Control, with a rf field, of photon emission times
  by a single molecule and its connection to laser-induced localization of an
  electron in a double well. \textit{J. Chem. Phys.} \textbf{2001,}
  \textsl{115,} 5989-5993.

\bibitem{osadko}
Osad'ko,~I.~S. Dynamic theory of two-photon correlators in the spectroscopy of
  single impurity centers. \textit{J.E.T.P.} \textbf{1998,} \textsl{86,}
  875--887.

\bibitem{wild2}
Verberk,~R.;\ \ Orrit,~M. Photon statistics in the fluorescence of single
  molecules and nanocrystals: correlation functions versus distributions of on-
  and off-times. \textit{J. Chem. Phys.} \textbf{203,} \textsl{119,}
  2214--2222.

\bibitem{barkrev}
Jung,~Y.;\ \ Barkai,~E.;\ \ Silbey,~R.~J. A Stochastic Theory of Single
  Molecule Spectroscopy. \textit{Advances in Chemical Physics} \textbf{2002,}
  \textsl{123,} 199--266.

\bibitem{zheng1}
Zheng,~Y.;\ \ Brown,~F. L.~H. Single-Molecule Photon Counting Statistics via
  Generalized Optical Bloch Equations. \textit{Phys. Rev. Lett.} \textbf{2003,}
  \textsl{90,} Art. No. 238305.

\bibitem{zheng2}
Zheng,~Y.;\ \ Brown,~F. L.~H. Photon emission from driven single molecules.
  \textit{J. Chem. Phys.} \textbf{2003,} \textsl{119,} 11814--11828.

\bibitem{zheng3}
Zheng,~Y.;\ \ Brown,~F. L.~H. Single Molecule Photon Emission Statistics for
  Non-Markovian Blinking Models. \textit{J. Chem. Phys.} \textbf{2004,}
  \textsl{121,} 3238--3252.

\bibitem{zheng4}
Zheng,~Y.;\ \ Brown,~F. L.~H. Single molecule photon emission statistics in the
  slow modulation limit. \textit{J. Chem. Phys.} \textbf{2004,} \textsl{121,}
  7914--7925.

\bibitem{cook}
Cook,~R.~J. Photon number statistics in resonance fluorescence. \textit{Phys.
  Rev. A} \textbf{1981,} \textsl{23,} 1243--1250.

\bibitem{muk1}
Mukamel,~S. Photon statistics: Nonlinear spectroscopy of single quantum
  systems. \textit{Phys. Rev. A} \textbf{2003,} \textsl{68,} Art. No. 063821.

\bibitem{bark1}
He,~Y.;\ \ Barkai,~E. Influence of spectral diffusion on single-molecule photon
  statistics. \textit{Phys. Rev. Lett.} \textbf{2004,} \textsl{93,} Art. No.
  068302.

\bibitem{bark2}
He,~Y.;\ \ Barkai,~E. Super- and Sub-Poissonian photon statistics for single
  molecule spectroscopy. \textit{J. Chem. Phys.} \textbf{2005,} \textsl{122,}
  Art. No. 184703.

\bibitem{muk2}
Sanda,~F.;\ \ Mukamel,~S. Liouville-space pathways for spectral diffusion in
  photon statistics from single molecules. \textit{Phys. Rev. A} \textbf{2005,}
  \textsl{71,} Art. No. 033807.

\bibitem{loudon}
Loudon,~R. \textit{The Quantum Theory of Light;} Oxford: New York, third ed.;
  2000.

\bibitem{ct}
Cohen-Tannoudji,~C.;\ \ Dupont-Roc,~J.;\ \ Grynberg,~G. \textit{Atom-Photon
  Interactions;} Wiley-Interscience: New York, 1992.

\bibitem{mukbook}
Mukamel,~S. \textit{Principles of Nonlinear Optical Spectroscopy;} Oxford: New
  York, 1995.

\bibitem{blum}
Blum,~K. \textit{Density Matrix Theory and Applications;} Plenum Press: New
  York, second ed.; 1981.

\bibitem{schatz}
Schatz,~G.;\ \ Ratner,~M. \textit{Quantum Mechanics in Chemistry;} Prentice
  Hall: Englewood Cliffs, New Jersey, 1993.

\bibitem{vankampen}
van Kampen,~N.~G. \textit{Stochastic Processes in Physics and Chemistry;}
  North-Holland: Amsterdam, 1992.

\bibitem{optlett}
Mandel,~L. Sub-Poissonian photon statistics in resonance fluorescence.
  \textit{Opt. Lett.} \textbf{1979,} \textsl{4,} 205--207.

\bibitem{phillips}
Phillips,~W.~A. \textit{J. Low Temp. Phys.} \textbf{1972,} \textsl{7,} 351.

\bibitem{anderson}
Anderson,~P.~W.;\ \ Halperin,~B.~I.;\ \ Varma,~C.~M. Anomalous low-temperature
  thermal properties of glasses and spin glasses. \textit{Philos. Mag.}
  \textbf{1972,} \textsl{25,} 1--9.

\bibitem{silb}
Silbey,~R.  RELAXATION THEORY APPLIED TO SCATTERING OF EXCITATIONS AND OPTICAL
  TRANSITIONS IN CRYSTALS AND SOLIDS.   In  \textit{Relaxation Processes in
  Molecular Excited States}; Funfschilling,~J.,\ \ Ed.;  Kluwer Academic
  Publishers: 1989.

\bibitem{brownsilbey}
Brown,~F. L.~H.;\ \ Silbey,~R.~J. An investigation of the effects of two level
  system coupling on single molecule lineshapes in low temperature glasses.
  \textit{J. Chem. Phys.} \textbf{1998,} \textsl{108,} 7434--7450.

\bibitem{kubo}
Kubo,~R.  A stochastic theory of line-shape and relaxation.   In
  \textit{Fluctuation, Relaxation and Resonance in Magnetic Systems};
  TerHaar,~D.,\ \ Ed.;  Oliver and Boyd: Edinburgh, 1962.

\bibitem{frank_rev}
Brown,~F. L.~H. Generating function methods in single molecule spectroscopy.
  \textit{Accounts of Chemical Research} \textbf{2006,}  In press.

\bibitem{friesner}
Pollard,~W.~T.;\ \ Friesner,~R.~A. Solution of the Redfield equation for the
  dissipative quantum dynamics of multilevel systems. \textit{J. Chem. Phys.}
  \textbf{1994,} \textsl{100,} 5054--5065.

\end{thebibliography}
\providecommand{\refin}[1]{\\ \textbf{Referenced in:} #1}

\end{document}